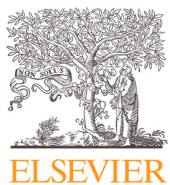
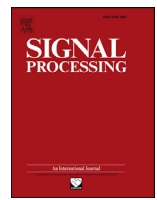

# Decomposition of higher-order spectra for blind multiple-input deconvolution, pattern identification and separation

Christopher K. Kovach*, Matthew A. Howard III

*Department of Neurosurgery, University of Iowa Carver College of Medicine, Iowa City, Iowa, United States*

**A B S T R A C T**

Like the ordinary power spectrum, higher-order spectra (HOS) describe signal properties that are invariant under translations in time. Unlike the power spectrum, HOS retain phase information from which details of the signal waveform can be recovered. Here we consider the problem of identifying multiple unknown transient waveforms which recur within an ensemble of records at mutually random delays. We develop a new technique for recovering filters from HOS whose performance in waveform detection approaches that of an optimal matched filter, requiring no prior information about the waveforms. Unlike previous techniques of signal identification through HOS, the method applies equally well to signals with deterministic and non-deterministic HOS. In the non-deterministic case, it yields an additive decomposition, introducing a new approach to the separation of component processes within non-Gaussian signals having non-deterministic higher moments. We show a close relationship to minimum-entropy blind deconvolution (MED), which the present technique improves upon by avoiding the need for numerical optimization, while requiring only numerically stable operations of time shift, element-wise multiplication and averaging, making it particularly suited for real-time applications. The application of HOS decomposition to real-world signals is demonstrated with blind denoising, detection and classification of normal and abnormal heartbeats in electrocardiograms.

## 1. Introduction

Deconvolution arises, implicitly or explicitly, in many signal-processing problems, ranging from compression and denoising to system identification. It has particular relevance to broadly applicable detection problems, as in finding the time at which some fixed recurring, known or unknown, signal of interest appears in a background of stationary noise. In the case of both known and unknown (or imperfectly known) signals, the solution is typically a matter of designing a filter that yields an impulsive output from which the timing of the target signal may be inferred [66], essentially a problem of deconvolution. In the case of an unknown signal, the problem is that of *blind* deconvolution. Methods of blind deconvolution commonly identify filters that optimize some measure of sparseness, such as kurtosis, producing an output whose energy is focused at a small number of impulse-like peaks. For this reason, problems of signal detection and delay estimation provide a useful and easily grasped point of entry for thinking more generally about blind deconvolution.

When an unknown signal is present with additive noise and minimal phase distortion in multiple sensors or records, one might try to identify the signal by first working out the delays between the respective input records, after which the form of the signal might be recovered through simple averaging. One could try to estimate delays using second order techniques, which essentially take the available noise-corrupted instances of the signal as stand-ins for optimally matched filters [12,68]. But to work, such techniques require consistent estimators of the relevant second-order statistics, which are often lacking due to second-order non-stationarity. Examples to which this applies include the identification of a transient signal at a low signal-to-noise ratio within an array of sensors for which the variability of arrival times is large relative to the signal duration. Similar scenarios arise in the identification of recurring patterns through the comparison of sequentially recorded epochs from a longer record, making the problem relevant for pattern identification.

A chief aim of the present work is to show how information in *higher-order spectra* (HOS), which are Fourier domain represen-tations of signal higher-order moments [8,20,47,49], apply to the problem of detecting an unknown signal as well as towards an understanding of what moment-based techniques of deconvolution accomplish. In pursuing this goal, we identify a method for separating additive independent signals through a decomposition of HOS (HOSD). Promising aspects of this approach include the fact that all of the steps in the proposed algorithm involve computationally simple and numerically stable operations, time shifting, averaging and element-wise multiplication, which are readily adapted to real-time applications.

* Corresponding author.
*E-mail address:* christopher-kovach@uiowa.edu (C.K. Kovach).

*1.1. Organization*

Higher-order spectra have a longstanding reputation for resisting simple and intuitive interpretation [8]. We have tried to counter this tendency by emphasizing an intuition-friendly framework of ideas over rigor and detail, leaving much fleshing-out for the future. Accordingly, we focus the development on univariate real-valued signals, treating the generalization to complex and multivariate signals as supplemental to the main development, within Sections 2.1.5, 2.1.6, and Appendix A. We have also favored the continuous Fourier transform over the Z-transform, moving somewhat cavalierly from the continuous case to examples using finitely sampled time series. The main ideas are built up from easily grasped applications to detection and delay estimation, which serve as platforms for a more generally applicable exploration of the topic.

In keeping with the tutorial aim, Section 2.1 provides a descriptive review of the main relevant properties of HOS, followed in Section 2.1.4 by a novel and, it is hoped, simpler treatment of the relationship between cumulant and moment HOS in terms of filter-like window functions applied to HOS. Several preliminary results used later to show the relationship between the present technique and deconvolution through moment maximization are given in Section 2.2. Section 3 develops the general strategy for obtaining filters optimized for detection using arguments similar to those of matched filter theory [66], with little mention of HOS at first. From there we progress to the development of detection filters from HOS in Section 4; their application to non-deterministic signals in Section 5; and to signal estimation and recovery in Section 6. The many points of contact between the present technique and classical techniques of blind deconvolution through moment-maximization are reviewed in Section 7. Section 8.1 describes some implementations of the technique, with software for Matlab available for download [35]. Finally, Section 10 demonstrates the performance of the algorithm with artificial test signals and real ECG data, alongside comparisons to alternative blind and non-blind techniques.

*1.2. Signal model*

We will be concerned with arrays of real-valued signal, $\{x_i(t)\}$, which contain one or more randomly recurring features, $f_j(t)$, subject to a random non-negative scaling, $a_{ij}$, and random delay, $\tau_{ij}$, in a background of independent additive colored Gaussian noise, without phase distortion:

$$x_i(t) = \sum_{j=1}^{P} a_{ij} f_j(t - \tau_{ij}) + n_i(t) \quad (1)$$

It will be assumed that the feature occurrence times, $\tau_{ij}$'s, are mutually independent across records and feature-generating processes, and that all relevant moments of the signal are finite and identifiable. The separate $x_i$'s need not be concurrent records; in example applications to ECG later, they will correspond to sequentially recorded epochs from a single long-duration record. We suppose only that the records contain some common set of features, $f_j$'s, whose identification and recovery is our goal. Unless otherwise noted, it will be further assumed that noise processes are wide-sense stationary, that the relevant signal moments are bounded and observable, and that the signal is continuous, such that its autocorrelation is twice differentiable at zero lag.

## 2. Background

*2.1. Review of higher-order spectra*

The most important property of HOS, for our purposes, is that they describe forms of statistical dependence within a signal that are invariant under translation in time [8,47,49]. For a real-valued signal, $x(t)$, with Fourier transform, $X(\omega) = \mathcal{F}\{x(t)\}$, given a time shift, $\tau$, we have

$$\mathcal{F}\{x(t-\tau)\} = X(\omega)e^{-i\omega\tau} \quad (2)$$

where $\mathcal{F}$ denotes the continuous Fourier transform. HOS relate to products in the signal spectrum:

$$M^K[X] = \mathrm{E}[X(\omega_1)X(\omega_2)\ldots X(\omega_K)] \quad (3)$$

taken at combinations of frequencies which sum to zero: $\sum_{k=1}^{K} \omega_k = 0$. Such products are time-shift invariant because the exponential terms from (2) mutually cancel. Time-shift invariant products of $K$ frequencies constitute the $K$th order spectrum. Recalling that for a real-valued signal $X(-\omega) = X^*(\omega)$, the second-order spectrum is $\mathrm{E}[X(\omega)X(-\omega)] = \mathrm{E}[|X(\omega)|^2]$, which is clearly the same as the power spectrum. The third-order spectrum is also known as the bispectrum:

$$M^3(\omega_1, \omega_2) = \mathrm{E}[X(\omega_1)X(\omega_2)X^*(\omega_1 + \omega_2)] \quad (4)$$

and the fourth-order as the trispectrum:

$$M^4(\omega_1, \omega_2) = \mathrm{E}[X(\omega_1)X(\omega_2)X(\omega_3)X^*(\omega_1 + \omega_2 + \omega_3)] \quad (5)$$

Unlike the power spectrum, HOS do not discard all phase information, but are selective in discarding linear phase arising from time shifts. In fact, it may be shown that for transient signals, the signal phase spectrum is fully recoverable from its deterministic HOS [3,40,42,53,64]. For this reason, a signal waveform may be recovered from deterministic HOS without having to know where the signal appears in the observed time series. The possibility of recovering a signal in this way has attracted sporadic attention over multiple decades [1,3,6,32,40,42,48,53–55,64]; yet the range of practical applications arising from it has remained relatively circumscribed. We will briefly consider the reasons for the limited success of previous techniques later.

Finally, HOS may be used to cleanly separate non-Gaussian signal from Gaussian noise because any linear-time invariant Gaussian process is fully characterized by second-order statistics, precluding any form of cross-frequency dependence. It is therefore possible to recover HOS from randomly shifted signals in a background of Gaussian noise without noise-related bias. Moreover, if the signal is transient, meaning of finite duration, its energy must be infinitely extended over frequency and thus have non-vanishing HOS *at all orders*. This holds in principle—in practice, one cares only about the portion of the spectrum whose energy is sufficient to stand a chance of being recovered from noise, which might be narrow or, in some cases, non-existent if the transient has a highly oscillatory character, as explained next.

*2.1.1. Bandwidth selectivity*

In contrast to second-order techniques, HOS-related techniques, particularly those of low odd orders, apply most naturally to spectrally broad signal components. The signal spectrum is certain to be fully recoverable only for transient signals and partially recoverable for band-limited signals that meet certain bandwidth requirements. Specifically, for a band-limited signal with a high-pass frequency of $W_0$ to have a non-vanishing $K$th-order moment, it must also have a low-pass frequency, $W_1 \geq W_0$ such that

$$W_1 \geq \frac{p}{K-p}W_0 \qquad (6)$$

for some integer $p$, otherwise all moment products must include at least one term that lies outside the bandwidth of the signal, causing them to vanish. We see from this that HOS of even order have no minimum bandwidth (given that we may set $K = 2p$); but to have non-vanishing odd-order HOS, a signal must have a minimum low-pass frequency (setting $K = 2p - 1$):

$$W_1 \geq \frac{K+1}{K-1}W_0 \qquad (7)$$

corresponding to a fractional bandwidth of

$$2\frac{W_1 - W_0}{W_1 + W_0} \geq \frac{2}{K} \qquad (8)$$

The bispectrum, for which $K = 3$, is thus uniquely selective for spectrally broad components with fractional bandwidth greater than 2/3. Odd-order HOS of higher orders may accommodate signals with successively narrower bandwidths. In general, however, the narrower the bandwidth, the greater the ambiguity in the signal spectrum recovered from HOS.

### 2.1.2. Symmetry

The product in (3) clearly does not change under the $K!$ possible permutations of the frequency arguments, implying that the full spectrum can be recovered from any of multiple redundant symmetry regions. In addition to permutation, HOS of real-valued signals are conjugate symmetric under a reversal in the sign of all frequency arguments, doubling the symmetry; thus there are $2K!$ redundant regions (except when $K = 2$, for which permutation coincides with sign-reversal) [21]. This point is important in developing efficient estimators of HOS, as only one symmetry region needs to be considered for the purpose of estimation.

### 2.1.3. Deterministic vs. non-deterministic HOS

A signal with a fixed waveform, possibly subject to a random time shift and random non-negative scaling, will be referred to as *deterministic*. For deterministic HOS, the expectation (3) may be written as a product of $K$ terms within the spectrum of the signal. On the other hand, *non-deterministic* HOS cannot, in general, be written as such a product [41] but might be expanded as a sum of products, in analogy to the singular value decomposition of a covariance matrix. In particular, non-deterministic cumulant HOS resulting from multiple independent additive processes, each of which is separately deterministic, can be written as a sum over deterministic HOS in the same way that a covariance matrix can be written as a sum over outer products. As described shortly, the distinction between deterministic and non-deterministic HOS is critical for waveform recovery. We note that a few authors have reserved the term "bispectrum" for what is here referred to as the "deterministic bispectrum" [6,41].

### 2.1.4. Moment, cumulant and quasi-cumulant spectra

Cumulants differ from moments in being additive for independent additive processes and vanishing at higher orders for Gaussian processes. The additive property of cumulants makes them relevant for blind separation of independent processes [11]; but because they do not typically have the simple monomial form of moments, cumulants tend to be mathematically more cumbersome to handle than moments. In the spectral domain, the divergence between moment and cumulant is confined to subdomains in which a deterministic moment spectrum may be expressed as the product of lower-order moment spectra; for example, in the trispectrum, along $\omega_1 + \omega_2 = 0$ and $\omega_3 + \omega_4 = 0$, where the deterministic 4th-order moment spectrum reduces to the product of the power spectrum at different frequencies. For signals with deterministic HOS, the conversion from moment to cumulant amounts to windowing out corresponding regions of the spectrum, which can be done explicitly when working with HOS [9]. Such explicit windowing leads to families of measures that behave like cumulants. We will refer to these as *quasi-cumulants*.

**Definition 1** (Quasi-cumulant). Quasi-cumulants result from applying a window, $Q$ to moment HOS, $M^K$, such that

$$Q(\omega_1, \ldots, \omega_{K-1})M^K(\omega_1, \ldots, \omega_{K-1})$$
$$= 0 \text{ for any } \left\{\omega_k \mid \sum_{j=1}^{J} \omega_{k_j} = 0, J \leq K - 2\right\} \qquad (9)$$

and $Q(\omega_1, \ldots, \omega_{K-1}) \geq 0$ otherwise. Cumulant spectra belong to the family of quasi-cumulants only if the cumulant spectrum vanishes within the suppressed region, which happens under two conditions: (1) if the signal is deterministic up to a time shift and scaling, and (2) if a non-deterministic signal exhibits no dependence within the corresponding HOS domain.

In practice, one often obtains estimators by averaging over short intervals within longer records, which amounts to smoothing the deterministic HOS of the full record [9,37,61]. For the purpose of estimation, one might begin by applying quasi-cumulant windowing to the deterministic spectrum of the full record before computing the smoothed estimate [9], but it will often prove computationally expedient to compute the moment HOS before windowing, discarding information within a larger bandwidth in the moment HOS. Because HOS estimators are subject to broadband bias, this approach entails the suppression of intervals in the HOS domain:

$$Q(\omega_1, \ldots, \omega_{K-1})M^K(\omega_1, \ldots, \omega_{K-1})$$
$$= 0 \text{ for any } \left\{\omega_k \mid \sum_{j=1}^{J} \omega_{k_j} \leq \Delta W, J \leq K - 2\right\} \qquad (10)$$

where $\Delta W$ is the characteristic frequency resolution of the estimator. In developing optimal delay estimators, we will already consider weighting functions in the higher-order spectral domain, thus it will not add much complication to restrict attention to estimators that apply such quasi-cumulant weighting.

### 2.1.5. Higher-order cross-spectra and extensions to multivariate signals

Higher-order cross-spectra are obtained by drawing terms in the HOS moment product (3) from two or more elements of a multivariate signal. They generalize the ordinary cross-spectrum in the same way that higher-order auto-spectra generalize the power spectrum, yielding measures of multi-way spectral dependence. The complete set of $K$th-order HOS within a multivariate signal with $m$ components contains all $m^K$ spectra arising in the expansion of $(X_1 + X_2 + \cdots + X_m)^K$, of which $\binom{K+m-1}{m-1}$ are non-redundant:

$$M^K[X_1 + X_2 + \cdots + X_m]$$
$$= \sum_{\{j_1, \ldots, j_k\}} E\big[X_{j_1}(\omega_1)X_{j_2}(\omega_2)\ldots X_{j_k}(\omega_K)\big]\delta\left(\sum_{k=1}^{K}\omega_k\right)$$
$$= \sum_{\{j_1, \ldots, j_k\}} M_{\{j_1, \ldots, j_k\}}[X] \qquad (11)$$

where $\delta$ notation is used to indicate the restriction to $\sum_{k=1}^{K}\omega_k = 0$ and $j_k \in \{1, 2, \ldots, m\}$. Each non-redundant spectrum in this expansion is symmetric under permutation of the arguments across the $p_j$ terms drawn from the $j$th signal component, giving altogether $2\prod_{j=1}^{m} p_j!$ symmetry regions, where $\sum_{j=1}^{m} p_j = K$. Thus, there are $\frac{2K!}{2\prod_{j=1}^{m} p_j!} = \binom{K}{p_1,\ldots,p_m}$ non-redundant regions in the spectrum separately and $m^K$, in total across spectra, needed to describe the complete set of $K$th-order auto- and cross-interactions of an $m$-component signal.

#### 2.1.6. Extension to complex-valued signals

In considering how to interpret the HOS of a univariate complex-valued signal, $y(t) \in \mathbb{C}^1$, it will be useful to note an isomorphism with the HOS of a bivariate real signal, $x(t) \in \mathbb{R}^2$. But first, the definition given in (3) must be amended, as taking the complex conjugate of the signal spectrum is no longer interchangeable with reversing the sign of the frequency argument, rather the two operations yield distinct sets of time-shift invariant statistics [2]. The complete set of $K$th-order statistics involve multiple distinct spectra, which arise from the following more general definition:

$$M^{\{i_1,\ldots,i_K\}}[Y] = E\left[Y^{\{i_1\}}(\omega_1)Y^{\{i_2\}}(\omega_2)\ldots Y^{\{i_K\}}(\omega_K)\right]\delta\left(\sum_{k=1}^{K} i_k\omega_k\right) \quad (12)$$

where $i_k \in \{-1, 1\}$ and $a^{\{-1\}} = a^*$ here denotes the complex conjugate and $a^{\{1\}} = a$ denotes identity. Because one can contrast any HOS product involving $Y(\omega_k)$ with a distinct product involving $Y^*(-\omega_k)$, there are $2^K$ additional products to consider.

The isomorphism between the HOS of univariate complex and bivariate real signals can be observed by decomposing $y$ into the analytic transforms of two real signals, $x_+(t) \in \mathbb{R}^1$ and $x_-(t) \in \mathbb{R}^1$, unique up to a constant offset, with support over negative and positive frequencies, respectively:

$$y = \frac{1}{2}(x_+ + x_-) + i\frac{1}{2}\mathscr{H}\{x_+ - x_-\} \quad (13)$$

for which

$$Y(\omega) = X_-(\omega)I_{(-\infty,0]}(\omega) + X_+(\omega)I_{[0,\infty)}(\omega) \quad (14)$$

Using the equivalence of complex conjugation and sign reversal in the frequency argument for the real signals, we have

$$M^{\{i_1,\ldots,i_K\}}[Y](\omega_1,\ldots,\omega_K)$$
$$= E[X_{j_1}(i_1\omega_1)X_{j_2}(i_2\omega_2)\ldots X_{j_K}(i_K\omega_K)]\delta\left(\sum_{k=1}^{K} i_k\omega_k\right)$$
$$\text{with} \quad j_k = \text{sgn}[\omega_k] \quad (15)$$

The isomorphism can be made explicit with a change of variables and rearrangement of indexed values, such that $\omega'_k = i_k\omega_k$ and $j'_k = i_k$, using the fact that $\{i_k\} \cong \{j_k\}$. The mapping that results from this is illustrated for $M^{\{1,1,-1\}}$ in Fig. A.1.

It is evident from this that fully characterizing the HOS of an arbitrary complex signal requires information in $2^K$ symmetry regions across all component HOS, in contrast to the single region required for real-valued signals. However, in the special case when $y$ is analytic, such that $x_- = 0$, the HOS of $y$ is clearly isomorphic to that of $x_+ = \text{Re}\{y\}$, and only one symmetry region needs to be considered, as in the real-valued case.

For the same reason that cross-spectra appear within the multinomial expansion of the sum of components in a multivariate signal, the moment of the real part of $y$ contains a summation over the component spectra of a complex signal; that is to say

$$M^K[\text{Re}\{y\}] = M^K[Y+Y^*] = \sum_{\{i_1,\ldots,i_K\}} M^{\{i_1,\ldots,i_K\}}[Y]$$
$$= M^K[X_+ + X_-] = \sum_{\{j_1,\ldots,j_K\}} M^{\{j_1,\ldots,j_K\}}[\{X_+, X_-\}] \quad (16)$$

This point is highly relevant for adapting moment-maximizing deconvolution techniques to complex signals, as discussed shortly.

In the interest of simplicity, the emphasis here will remain on real-valued univariate signals, but the preceding discussion shows that extensions to multivariate and complex signals are straightforward, with a few important subtleties. These are considered further in Appendix A. The main practical differences between applications to univariate and multivariate signals arise from the need to account for additional symmetry regions within higher-order cross-spectra. The extension to multivariate complex signals is likewise straightforward: the HOS of a complex $m$-variate signal is isomorphic to that of a $2m$-variate real signal whose components are derived from the separate positive and negative frequency domains of respective components, as in the univariate case.

As a final note, it is quite common to find higher-order spectra or their time-domain equivalents and related zero-lag measures applied to complex signals in ways that do not encompass a complete set of non-redundant symmetry regions, thus neglecting some aspects of the $K$th-order statistics of a general complex-valued process [3,6,15,18,19,26,29,63–65]. This is sometimes justified by the nature of the signal (e.g. $y$ is analytic) or the application, but because the scope of a given application is not always made explicit, the literature should be read with caution on this point. In the context of HOS inversion [3,6], any spectrum for which $\left|\sum_{k=1}^{K} i_k\right| \in \{1, 2\}$ contains enough information to uniquely recover a complex deterministic signal (see Corollary 1.5 in the following). In practice, however, differences in the sampling of symmetry regions imply that the component spectra do not all give precisely the same information about the signal. In the context of zero-lag moment maximization [15,19,63,65], Eq. (16) shows that the most general objective function, which sums over all component spectra, is simply the zero-lag moment of the real part of the filtered signal, $\text{Re}\{h^*y\}$, a point that appears to have received little attention in the literature.

#### 2.1.7. Complex modulation and HOS

Many applications involve signals that are shifted in the frequency domain such that signal statistics must be computed against a reference frequency, other than $\omega = 0$, or some other carrier waveform. While second-order spectra are unaltered, beyond frequency translation [36], narrowband carrier signals do not similarly preserve HOS, and recovery of the original statistical structure requires explicit demodulated first. While modulation by a narrowband carrier signal does not generalize to HOS, other modulation schemes may. In particular, a carrier whose HOS contains an impulse effects a translation in the HOS domain; such a carrier is composed of K complex sinusoids whose frequencies sum to zero.

### 2.2. HOS and moment-maximizing filters

In establishing the relationship between HOS decomposition and moment maximization, we will frame the latter as the search for a projection in the space of HOS. The following lemma is useful for understanding the link between moment maximization and signal recovery.

**Lemma 1** (Matched filters maximize all zero-lag moments). *For a real-valued signal, $x(t) \in \mathbb{R}^1$, deterministic up to an arbitrarily large time shift and non-negative scaling and having a finite non-zero $K$th-order moment, the matched filter,*

$$h(t) = \frac{x(-t)}{\left[\int (x \star x)^K(t)\,dt\right]^{\frac{1}{2K}}} \quad (17)$$

*maximizes the zero-lag $K$th-order moment among filters of unit energy in the $K$th-order HOS; that is, with $\int (h \star h)^K(t)\,dt = 1$, where $\star$ denotes cross-correlation.*

**Proof.** Because the $K$th-order HOS is the Fourier transform of the $K$th-order autocorrelation, the zero-lag $K$th-order moment of the filtered signal is given by the integral over the corresponding HOS:

$$\int (h*x)^K(t)\,dt = \int \ldots \int H(\omega_1)X(\omega_1) \\ \ldots H(\omega_K)X(\omega_K)\delta(\omega_1+\cdots+\omega_K)\,d\omega_1\ldots\omega_K \quad (18)$$

We may regard (18) as an inner product between the deterministic HOS of X and that of the complex conjugate of H within an inner product space that is a superset of all $K$th-order HOS [46]. By a direct application of the Cauchy–Shwartz inequality, we have

$$\int (h*x)^K(t)\,dt \leq \sqrt{\int \int (x\star x)^K(t)\,dt} \quad (19)$$

with equality for $h$ in Eq. (17). □

**Corollary 1.1** (Lemma 1 applies to cumulants and quasi-cumulants). *The matched filter maximizes all zero-lag (quasi-)cumulants of x among filters with unit energy in the corresponding (quasi-)cumulant HOS.*

**Proof.** Under quasi-cumulant windowing of HOS as in (9), a zero-lag quasi-cumulant of $h*x$ yields a semi-inner-product within the space of moment spectra, which weights each term in the integrand with a non-negative value. The Cauchy–Schwartz inequality therefore applies in the same way as before. The argument extends to zero-lag cumulants of signals with deterministic HOS because their cumulant HOS may be obtained with quasi-cumulant windowing. □

**Corollary 1.2** (Lemma 1 applies to the phase response separately from amplitude response). *For a filter, h, of a given amplitude response, zero-lag quasi-cumulants and moments are maximized when the phase response of h is matched to the phase spectrum of x.*

**Proof.** When the phase response of $h$ is matched to $x$, all terms in the integrand of (18) have a common phase of zero, being real, non-negative and equal to magnitude. By the triangle inequality applied to the summation of phasors, the integral must therefore attain a maximum. □

**Corollary 1.3** (Uniqueness of the maximizing filter for a transient signal). *The filter with phase response matched to the phase spectrum of a transient and deterministic x (up to time shift and scaling) is a unique maximizer of the Kth-order moment and cumulant if $K > 2$, among filters with fixed amplitude response.*

**Proof.** We have established that the $K$th-order HOS of the maximizing filter function must be matched to that of $x$, so that the HOS of the filtered signal is uniformly real and non-negative:

$$\sum_{k=1}^{K}[\phi_x(\omega_k)+\phi_h(\omega_k)] = 2\pi\nu(\omega_1,\ldots\omega_{k-1}) \quad (20)$$

where $\nu(\omega_1,\ldots\omega_{k-1})$ assumes only integer values and $\omega_K = -\sum_{k=1}^{K-1}\omega_k$. Now we show that this requires the phase response of the filter, itself, to be matched to the phase spectrum of $x$, up to a time shift. Because $x$ has finite duration, its spectrum must have infinite support and its phase spectrum must be continuously differentiable modulo $2\pi$ (for the same reason that a band-limited signal has infinite duration and is continuously differentiable in time), and we may therefore ignore the possibility of solutions with non-trivial discontinuities. Taking the derivative of (20) and using $\frac{d\omega_K}{d\omega_j} = -1$

$$\frac{\partial}{\partial \omega_j}[\phi_x(\omega_j)+\phi_h(\omega_j)] - \frac{\partial}{\partial \omega_K}[\phi_x(\omega_K)+\phi_h(\omega_K)] = 0 \quad (21)$$

The phase derivative (group delay) may therefore, at most, differ between $x$ and $h$ by some constant, so that

$$\phi_h(\omega) = -\phi_x(\omega) + b\omega + c \quad (22)$$

Thus, by (20)

$$cK \mod 2\pi = 0$$

and $\phi_h$ and $\phi_x$ can differ at most by a constant group delay for real-valued $x$, corresponding to a time shift, and by sign, for even orders. That this argument must apply as well to cumulant spectra, everywhere but $\phi(0)$, follows from the fact that cumulant HOS differ from moment HOS only in subdomains of infinitesimal support, which exclude only $X(0)$ entirely from the cumulant HOS. □

**Corollary 1.4** (The output of a maximizing filter is impulsive). *The moment- and cumulant-maximizing filter, h, for a deterministic transient signal, x, with non-vanishing HOS produces an impulsive output, meaning here an output characterized by a symmetric zero-phase function centered at a lag time.*

**Proof.** This follows from the fact that the unique maximizing filter is matched to the phase spectrum of $x$. □

**Corollary 1.5** (Extension to complex signals). *Lemma 1 applies to a complex univariate deterministic signal, $y(t) \in \mathbb{C}^1$.*

**Proof.** To extend the preceding argument to complex-valued signals, recall that multiple component spectra must be considered, according to (12). The zero-lag moment obtained by substituting Re$\{h*y\}$ for $h*x$ in (18) is a natural choice for an objective function, because it sums over all component spectra as well as frequency dimensions.

$$\int (\text{Re}\{h*y\})^K(t)\,dt = \sum_{\{i_1,\ldots,i_k\}} \int \ldots \int M_{\{i_1,\ldots,i_k\}}[HY]\,d\omega_1\ldots\omega_K \quad (23)$$

Eq. (23) is maximized when the HOS of the filtered signal is uniformly non-negative across frequencies as well as spectra, implying

$$\sum_{k=1}^{K} i_k[\phi_y(\omega_k)+\phi_h(\omega_k)] = 2\pi\nu(\omega_1,\ldots\omega_{k-1}) \quad (24)$$

for all combinations of $\{i_k\}$, where $i_k \in \{-1,1\}$. For each spectrum, $\frac{d\omega_K}{d\omega_j} = -i_j i_K$, and it can be verified that Eq. (22) still follows in each case. In addition, $c\sum_{k=1}^{K} i_k = (0 \mod 2\pi)$, leading to the following $K+1$ (redundant) constraints arising from the respective component HOS:

$$(2n-K)c = (0 \mod 2\pi) \text{ for } n = \{0,1,\ldots,K\} \quad (25)$$

For even $K$, the constraints arising from $n = 1 \pm K/2$ require $c = 0 \mod \pi$, while for odd $K$, $n = (1 \pm K)/2$ require $c = 0 \mod 2\pi$. These results show that $h$ is ambiguous only with respect to sign and only for even $K$, as we observed for $x(t) \in \mathbb{R}^1$. They imply, further, that the spectrum of a deterministic transient signal can be recovered from any $M^{\{i_1,i_2,i_3\}}[y]$ in isolation, provided $\sum_{k=1}^{K} i_k = \pm 2$, for even $K$, or $\sum_{k=1}^{K} i_k = \pm 1$ for odd $K$. □

### 2.3. Relationship to previous work

Multiple authors have considered the explicit application of HOS to signal identification, detection and delay estimation. Previous techniques have, however, dealt almost exclusively with deterministic HOS, [1,3,6,32,40,42,48,53–55,60,64], as has previous work applying HOS [30,50] and higher-order cross-correlations [56–58] to delay estimation. The deterministic assumption simplifies signal identification with HOS; for example, it allows the waveform to be recovered by log transforming the estimate and applying cepstral and related techniques towards the recovery of phase and magnitude spectra. Such techniques are not appropriate as a treatment of non-deterministic HOS, and the present approach dispenses with this assumption, seeking instead to decompose non-deterministic HOS into an additive series.

A decomposition of HOS according to distinct deterministic features may be regarded as implicit in convolutional independent component analysis (ICA), particularly in popular variants of ICA that rely on moment maximization [5,14,31], related applications of minimum entropy deconvolution (MED) to MIMO source recovery [15,19,63,65], and in the application of tensor decomposition techniques to higher moments [7,23,62]. Of the most direct relevance to the present development are applications of blind cumulant-based deconvolution to the recovery of MIMO sources [15,65] which estimate component signals by treating the result of deconvolution as a proxy for a known input. Similarities and differences between the present approach and these techniques will be considered.

Finally, some precedent for the linear decomposition of HOS exists in the literature on pattern classification, where higher-order correlations are occasionally used as inputs to classifiers [38,46,59]. In this setting, higher-order correlations are usually treated as generic feature vectors. Any resulting basis set therefore has no particular relationship to the distinction between deterministic and non-deterministic HOS. The present work may therefore be regarded as following on several bodies of precedent: the use of HOS in delay estimation, waveform recovery, pattern classification, decompositions of HOS implicit in techniques of ICA and blind deconvolution.

## 3. Scheme for filter estimation

The following section introduces a simple framework that will be used to find optimal filters for detection and delay estimation, applicable to HOS of any order. It will be developed first for traditional second-order delay estimators in order to establish that it yields the expected optimal estimators in the familiar setting. In the case of a known signal, it will recover the matched filter [66], while in estimating the delay between two noisy signals, it yields the coherence-derived maximum-likelihood delay estimator [12]. The development for HOS follows in the subsequent section.

### 3.1. Optimal detection filters

To begin, we seek a linear FIR filter that, given the observed signal, $x$, containing some feature, $f$, embedded at delay, $\tau$, yields an expected output containing a peak at $\tau$. As a criterion of optimization, one might use the ratio of expected squared peak magnitude to noise variance [66]. Here, instead, we use the minimization of the expected variance of the peak location under perturbation by wide-sense stationary noise (WSS), for reasons explained below. It will be shown that solving this problem is a matter of finding the filter that maximizes the ratio of second derivatives at zero lag in the respective autocorrelations of the filtered signal and noise. The main restriction is therefore that the signal autocorrelation, $f^{(2)} = f \star f$, is twice differentiable at zero lag, which means that $f$, itself, must be continuous.[1]

Suppose that $f$ is embedded in a background of independent stationary Gaussian colored noise with spectrum $E[|N(\omega)|^2] = \sigma(\omega)$:

$$x(t) = f(t - \tau) + n(t) \qquad (26)$$

We wish to identify the lag at which $f$ appears in $x$ using a filter, $h$, whose output should contain a peak as close to $\tau$ as possible.

Consider the output,

$$r(t) = (h * x)(t) = (h * f)(t - \tau) + (h * n)(t) \qquad (27)$$

we require that the expectation of $r$ contains an extremum at $\tau$, meaning that

$$E[r'(\tau)] = E\left[\left.\frac{d}{dt}\right|_{t=\tau} r(t)\right] = 0 \qquad (28)$$

We also wish to minimize any perturbation of the extremum from $\tau$ by noise. A figure of merit for this purpose is the ratio between the expected squared first derivative, $E[(r'(\tau))^2]$, and the expectation of the second derivative, $E[r''(\tau)]$, both evaluated at $\tau$. This choice is justified with a second-order Taylor series expansion about $\tau$: the effect of any additive noise will be to shift the peak at $\tau$ by $\epsilon$, to a point at which the slope of the noiseless peak, in the second-order approximation, $E[r''(\tau)]\epsilon$, cancels with the slope of the observed signal (signal plus noise realization); that is

$$E[r''(\tau)]\epsilon \approx -r'(\tau) \qquad (29)$$

We may therefore design $h$ to minimize the squared magnitude of the expected perturbation, $E[\epsilon^2]$:

$$\rho = E[\epsilon^2] = \frac{E[(r'(\tau))^2]}{E[r''(\tau)]^2} \qquad (30)$$

The value of $\rho$ describes the variance of the peak location under the second-order approximation. It can therefore be taken as the variance of the delay estimator in the limit of high signal-to-noise ratio.

The Fourier transform of the expected second derivative of the filtered signal is

$$\mathcal{F}\{E[r''(t)]\} = E[\mathcal{F}\{r''(t)\}] = -\omega^2 E[X(\omega)] H(\omega)$$
$$= -\omega^2 F(\omega) H(\omega) e^{-i\omega\tau} \qquad (31)$$

hence, evaluated at $\tau$:

$$E[r''(\tau)] = \int -\omega^2 F(\omega) H(\omega) \, d\omega \qquad (32)$$

Similarly, expanding the square of the first derivative in the Fourier domain,

$$E\left[(r'(\tau))^2\right]$$
$$= E\left[\iint -\omega\xi \left(F(\omega) + N(\omega)e^{i\omega\tau}\right)\left(F(\xi) + N(\xi)e^{i\xi\tau}\right)\right.$$
$$\times H(\omega) H(\xi) \, d\omega \, d\xi \qquad (33)$$

Due to the fact that the noiseless term in the expansion of the integrand is the squared expected derivative at $\tau$, which vanishes, and the noise is assumed stationary and zero-mean, so that $E[N(\omega)] = 0$ and $E[N(\omega)N(\xi)] = \sigma(\omega)\delta(\omega + \xi)$, this reduces to

$$E\left[(r'(\tau))^2\right] = \int \omega^2 \sigma(\omega) |H(\omega)|^2 \, d\omega \qquad (34)$$

In other words, the expected squared slope of $r$ at $\tau$ is the negative of the second derivative of the autocorrelation of the filtered noise at 0 lag. Now we are prepared to find the filter, $h$, that optimizes the figure of merit from Eq. (30), $\rho$. This might be done by identifying where the derivative of Eq. (30) with respect to $|H(\omega)|$ vanishes, but the problem may be further simplified into an easily solved quadratic form by way of an equivalent Lagrangian function:

$$\Lambda = E\left[(r'(\tau))^2\right] + \rho\left(1 + E[r''(\tau)]\right) \qquad (35)$$

for which the extremum is found directly by completing the square:

---

[1] Technically speaking, "almost everywhere:" any discontinuity must contribute no measurable energy to the signal.

$$\Lambda = \int \omega^2 \sigma(\omega) \left| H(\omega) - \rho \frac{F^*(\omega)}{\sigma(\omega)} \right|^2 d\omega - \rho^2 \int \omega^2 \frac{|F(\omega)|^2}{\sigma(\omega)} d\omega + \rho \quad (36)$$

minimized at

$$H(\omega) = \rho \frac{F^*(\omega)}{\sigma(\omega)} \quad \text{for} \quad \omega \neq 0 \quad (37)$$

Because arbitrary rescaling of $H$ does not affect the result, the constant $\rho$ will be dropped henceforth.

This result is the same as the well-known matched filter [66]. At the frequency origin, $H$ is unconstrained, for which reason we may allow $H(0)$ to vanish:

$$H(\omega) = \rho \frac{F^*(\omega)}{\sigma(\omega)} I_{\omega \neq 0}(\omega) \quad (38)$$

which corresponds to the intuitive fact that any DC offset cannot affect the peak location and is therefore irrelevant. By setting $H(0) = 0$, we obtain a weighted spectrum that meets the definition of a quasi-cumulant, as given in Eq. (9). It will be left implicit that $H(0) = 0$ henceforth.

A slightly different signal model, which applies to the Wiener filter [39], leads to a form that combines signal and noise power in the denominator. The discrepancy arises because the model assumes both the noise and signal generating process to be stationary, whereas the matched filter assumes only a single instance of the signal within a given observation window. The stationary assumption is more appropriate if, for example, the occurrence of $f$ is driven by a stationary Poisson process; in such cases, the background of "noise" which perturbs the peak at $\tau$ includes any nearby $f$'s overlapping with the instance at $\tau$. Put differently, we want the estimator to optimally separate neighboring peaks from each other, as well as from the noise. These random instances of $f$ at delays other than $\tau$ contribute to background spectrum according to the power spectrum of the rate function of the emitting point process, $\lambda(\omega)$, combined with that of the feature, giving:

$$H(\omega) = \frac{F^*(\omega)}{\lambda(\omega)|F(\omega)|^2 + \sigma(\omega)} \quad (39)$$

If the point process is homogenous, $\lambda$ is equal to constant squared intensity [4]. The same argument applies for any WSS noise, whether Gaussian or not; the denominator always contains the combined total power of the signal and all other additive processes contributing to the error in the peak location.

A compelling practical reason to favor (39) even when the stationary model is not appropriate is that the denominator may be estimated directly from the total power of the observed signal, without having to disentangle the contributions of noise and signal. The two models also converge as signal-to-noise ratio diminishes, which tends to limit the practical importance of the distinction. Formally, the use of (39) as a general-purpose estimator may be justified by noting that it places an upper bound on (33). The integrand in Eq. (33) clearly attains a non-negative maximum where $\xi = -\omega$, implying that

$$E\left[(r'(\tau))^2\right] \leq E\left[A \int \omega^2 |F(\omega) + N(\omega)e^{i\omega\tau}|^2 |H(\omega)|^2 d\omega\right]$$
$$= A \int \omega^2 \left[|F(\omega)|^2 + \sigma(\omega)\right] |H(\omega)|^2 d\omega \quad (40)$$

for some constant, $A$. We are merely using total power as an upper bound on noise-only power.

The value of $r(\tau)$ relates to the signal-to-noise ratio. The expected squared value of $r(\tau)$ is:

$$E\left[r^2(\tau)\right] = E[r(\tau)]^2 + \int \frac{|F(\omega)|^2}{\sigma} d\omega = E[r(\tau)]^2 + E[r(\tau)] \quad (41)$$

This result implies that the variance of $r(\tau)$ is equal to its value: $\text{Var}[r(\tau)] = E[r^2(\tau)] - E[r(\tau)]^2 = E[r(\tau)]$. The expected value of $r(\tau)$ therefore comes normalized, such that,

$$\frac{E[r(\tau)]^2}{\text{Var}[r(\tau)]} = E\left[r^2(\tau)\right] - E[r(\tau)]^2 = E[r(\tau)] \quad (42)$$

But the conditions under which the value of $r(\tau)$ can be taken as a literal measure of signal-to-noise ratio are limited to the case when the feature, $f$, exhibits no variability of amplitude. It remains possible to recover $f$ even when it is subject to some variability of scaling, but in such cases $E[r(\tau)]$ shrinks in relation to the amplitude variance, causing $r(\hat{\tau})$ to give an overly conservative signal-to-noise ratio estimate.

The expected squared deviation of the peak from $\tau$ (i.e. $\rho$), can also be determined by substituting $H$ back into the definition of $\rho$, giving

$$\rho = \left[\int \omega^2 \frac{|F(\omega)|^2}{\sigma(\omega)} d\omega\right]^{-1} \quad (43)$$

In other words, the second derivative at peak times within the filtered signal gives an estimate of the local uncertainty in the true peak location. This result will be used later to motivate thresholding for signal reconstruction, for which the thresholded output provides a reasonable approximation of the peak distribution.

### 3.1.1. Delay between two noisy signals

The foregoing analysis extends to the case when the goal is to identify the relative delays between two noisy channels, both of which contain the signal. Now we have

$$x_1(t) = f(t - \tau_1) + n_1(t)$$
$$x_2(t) = f(t - \tau_2) + n_2(t) \quad (44)$$

where $n_1$ and $n_2$ are generated by zero-mean noise processes with independent phase, so that $E[N_1(\omega)N_2(\omega)] = 0$, but possibly correlated power. We wish to optimize $h$ to provide a peak near $\tau_1 - \tau_2$ in

$$r_{12}(t) = \int X_1(\omega)X_2^*(\omega)H(\omega)e^{i\omega t}d\omega \quad (45)$$

The same line of argumentation as in the previous case may be applied here, and it will not be repeated in detail, to obtain

$$H(\omega) = \frac{|F(\omega)|^2}{|F(\omega)|^2(\sigma_1(\omega) + \sigma_2(\omega)) + E\left[|N_1(\omega)|^2|N_2(\omega)|^2\right]} \quad (46)$$

If the noise in either channel vanishes, we find that the noise-free channel equates to $F$, and the integrand of Eq. (45) then contains Eq. (37), as expected. If the noise processes are identical and Gaussian, (46) simplifies to

$$H(\omega) = \frac{|F(\omega)|^2/\sigma^2(\omega)}{2|F(\omega)|^2/\sigma(\omega) + 1} \quad (47)$$

which is the same as the well-known maximum-likelihood delay estimator for Gaussian signals [12].

We have obtained these estimators with comparatively few starting assumptions about the signal distribution, which simplifies the extension to non-Gaussian signals. We may, for example, apply Eq. (46) to a non-Gaussian noise process for which power is correlated across channels within frequency bands, with independent phase. In this example, the estimator in (46) adds an additional penalty to the weighting of spectral regions where the correlation of power is high, while rewarding regions where power is anti-correlated across channels. For example, if the signal always appears randomly without noise in one channel and with noise in the other, so that $E[|N_1(\omega)|^2|N_2(\omega)|^2] = 0 \neq \sigma_1(\omega)\sigma_2(\omega)$, the integrand of Eq. (45) still reduces to Eq. (37).

So far we have considered the delay between isolated signals. If the underlying detection problem involves the possibility of multiple overlapping occurrences of the signal, we are, again, better off with the upper bound that includes both signal and noise power in the denominator, as in (40). In this case we have

$$H(\omega) = \frac{|F(\omega)|^2}{\mathrm{E}\big[|F(\omega)+N_1(\omega)|^2|F(\omega)+N_2(\omega)|^2\big]} \qquad (48)$$

which, for Gaussian noise, becomes

$$H(\omega) = \frac{|F(\omega)|^2}{\big(|F(\omega)|^2+\sigma_1(\omega)\big)\big(|F(\omega)|^2+\sigma_2(\omega)\big)} \qquad (49)$$

### 3.1.2. Delays between multiple noisy records

We are most interested in the problem of finding mutual delays between multiple channels or records. In this setting, the denominator in (48) may be estimated by averaging observed power over records. The trouble with second-order estimators arises from the numerator. The conventional solution takes the magnitude of the pairwise cross-spectra as the relevant delay estimator for a given pair, in which case the outcome is equivalent to obtaining a delay estimator from *coherency* between $x_i$ and $x_j$, weighted by *coherence*; that is, the estimator will have the magnitude of squared coherence while preserving the phase of coherency [12]. But in the present setting, no consistent estimator of the pairwise cross-spectra between records is available. Simply averaging the pairwise magnitude cross-spectra across all channels does nothing to suppress noise. We need the delays to estimate the signal, but the signal to estimate the delays, and both may be too corrupted to be recovered separately. We next describe how to overcome this impasse with higher-order spectra.

## 4. Delay estimators from HOS

Because HOS are time-shift invariant and also vanish for Gaussian noise, it will be possible to develop statistically consistent estimators of both the numerator and the denominator of the optimized filter, and therefore to recover an asymptotically optimal or near-optimal detection filter when doing so is not possible with second-order statistics. We assume initially that the background noise is independent and Gaussian. Stationary non-Gaussian noise with non-vanishing HOS will be considered later as well; while it is not directly possible to distinguish signal from noise HOS in such applications, the prospect of decomposing HOS into a series of spectra promises some additional tools for coping with non-Gaussian noise. The following development will focus on the third-order (bispectral) variant of the algorithm (HOSD3), although extensions to HOS of arbitrary order will be noted along the way.

The cross-bispectrum takes the product as in (4) over two or more channels. We will start by considering its application to the two-channel problem from Eq. (44). Note that the expected phase of the cross-bispectrum, like that of the ordinary cross-spectrum, encodes the lag between channels, $\Delta\tau_{jk} = \tau_j - \tau_k$:

$$M_{122} = \mathrm{E}[X_1(\omega_1)X_2(\omega_2)X_2^*(\omega_1+\omega_2)]$$
$$= F(\omega_1)F(\omega_2)F^*(\omega_1+\omega_2)e^{-i\omega_1\Delta\tau_{12}} \qquad (50)$$

Denote the output of the $K$th-order HOS filter of the $j$th channel as $r_{j_1\cdots j_K}$ with $r_j(t) = h*x_j(t)$. Pursuing the same strategy as before, define a filter, $H$, now in the two-dimensional bispectral domain, such that

$$r_{122}(t) = \int X_1(\omega_1)e^{i\omega_1 t}\int X_2(\omega_2)X_2^*(\omega_1+\omega_2)H(\omega_1,\omega_2)\,d\omega_2\,d\omega_1 \qquad (51)$$

The expectation of the second derivative reduces to

$$\mathrm{E}\big[r''_{122}(\Delta\tau_{12})\big] = -\int \omega_1^2 e^{i\omega_1\Delta\tau_{12}}\int \mathrm{E}[X(\omega_1)X(\omega_2)X^*(\omega_1+\omega_2)]$$
$$\times H(\omega_1,\omega_2)\,d\omega_2\,d\omega_1$$
$$= -\int \omega_1^2 F(\omega_1)\int F(\omega_2)F^*(\omega_1+\omega_2)H(\omega_1,\omega_2)\,d\omega_2\,d\omega_1 \qquad (52)$$

and the expectation of the square of the first derivative is:

$$\mathrm{E}\bigg[\big(r'_{122}(\Delta\tau_{12})\big)^2\bigg] = -\mathrm{E}\bigg[\int \omega_1 e^{i\omega_1\Delta\tau_{12}}X_1(\omega_1)X_2(\omega_2)$$
$$\times X_2^*(\omega_1+\omega_2)H(\omega_1,\omega_2)\,d\omega_2\,d\omega_1$$
$$\times \int \xi_1 e^{i\xi_1\Delta\tau_{12}}X_1(\xi_1)X_2(\xi_2)X_2^*(\xi_1+\xi_2)H(\xi_1,\xi_2)\,d\xi_2\,d\xi_1\bigg] \qquad (53)$$

To avoid a proliferation of minor terms, we will consider the bound on the estimator in Eq. (53), following the example in (40)

$$\mathrm{E}\bigg[\big(r'_{122}(\Delta\tau_{12})\big)^2\bigg] \leq A^2\int \omega_1^2 \mathrm{E}\big[|X_1(\omega_1)|^2|X_2(\omega_2)|^2|X_2(\omega_1+\omega_2)|^2\big]$$
$$\times |H(\omega_1,\omega_2)|^2\,d\omega_2\,d\omega_1 \qquad (54)$$

In this case, substituting (52) and (54) into the Lagrangian (35), and completing the square gives

$$H(\omega_1,\omega_2) = \frac{F^*(\omega_1)F^*(\omega_2)F(\omega_1+\omega_2)}{\mathrm{E}\big[|X_1(\omega_1)|^2|X_2(\omega_2)|^2|X_2(\omega_1+\omega_2)|^2\big]} \quad \text{for} \quad \omega_1\neq 0 \qquad (55)$$

The development for HOS more generally proceeds in the same way with cross-polyspectra of the form

$$M_{122\ldots 2}^K = \mathrm{E}\bigg[X_1(\omega_1)X_2(\omega_2)\ldots X_2(\omega_{K-1})X_2^*\bigg(\sum_{k=1}^K \omega_k\bigg)\bigg]$$
$$= F(\omega_1)F(\omega_2)\ldots F(\omega_{k-1})F^*\bigg(\sum_{k=1}^K \omega_k\bigg)e^{-i\omega_1\Delta\tau_{12}} \qquad (56)$$

yielding solutions with the same form as (55).

### 4.1. Polycoherence weighting

We may now proceed to the multi-channel case with $L$ records at our disposal. Contrary to the situation in the second-order case, a suitably consistent estimator of the numerator in (55) is found in the auto-bispectrum (or auto-polyspectrum, generally) computed across records. Various options for how we might estimate the denominator in (55) relate to different definitions of what is commonly called *bicoherence* (in the case of the bispectrum, *polycoherence* generally). Regardless of which definition is used, $H$ can be understood as bicoherence with squared denominator; that is, $\hat{H} = \hat{M}^3/D$ where bicoherence is $\beta = \hat{M}^3/\sqrt{D}$. In general, both the numerator of $H$ and terms within the denominator will be estimated across records with the assumption that the noise processes are identically distributed.

#### 4.1.1. Stationary Gaussian noise

If the noise process is stationary, Gaussian and identically distributed across records, then by the independence over frequencies of stationary Gaussian processes, the terms within the expectation of the denominator in (55) separate into the product of signal-plus-noise power spectra.

$$D \rightarrow \big(|F(\omega_1)|^2+\sigma(\omega_1)\big)\big(|F(\omega_2)|^2+\sigma(\omega_2)\big)$$
$$\times \big(|F(\omega_1+\omega_2)|^2+\sigma(\omega_1+\omega_2)\big) \qquad (57)$$

In this case we find that *H* itself separates into the product of three terms (or *K* terms in general), meaning that weighting will produce an outcome equivalent to computing cross-bispectra after filtering each signal with

$$H(\omega) = \frac{F^*(\omega)}{|F(\omega)|^2 + \sigma(\omega)}$$

This is the very same matched filter we obtained in the second-order case, but here we will often have access to consistent estimators of the numerator in Eq. (55) when none are available for second-order estimators. It is also the case that the final outcome of the bispectral filter, $r_{ijj}$ in Eq. (51), is not linear, which will provide some important advantages later in developing an efficient procedure for bringing multiple records into alignment.

### 4.1.2. Wide sense stationary noise

We may relax the Gaussian assumption by estimating the products in the denominator of (55) directly. A form closely related to the most common definition of bicoherence [33] is obtained here under the assumption that power in the noise process is independent across records, but not across frequency bands, in which case the denominator of (55) separates into the product of the power spectrum at $\omega_1$ and the expected product of signal-plus-noise power at $\omega_1$ and $\omega_1 + \omega_2$, that is

$$D = \langle |X(\omega_1)|^2 \rangle \langle |X(\omega_2)|^2 |X(\omega_1 + \omega_2)|^2 \rangle \tag{58}$$

Bicoherence is commonly defined with the numerator $D = <|X(\omega_1)|^2 |X(\omega_2)|^2> <|X(\omega_1 + \omega_2)|^2>$, which amounts to a transformation of the axes in (58). The magnitude of this form is constrained to fall between 0 and 1, as expected for $F(\omega_1)F(\omega_2)F^*(\omega_1+\omega_2)H(\omega_1,\omega_2)$; for this reason, it is often favored as a normalized measure of cross-frequency dependence. A similar argument applies at higher orders as well.

### 4.1.3. Magnitude-weighted bicoherence

A variation less easy to justify from first principles, but having some practical advantages, normalizes by the square of the average magnitude:

$$D = \langle |X(\omega_1)X(\omega_2)X(\omega_1+\omega_2)| \rangle^2 \tag{59}$$

Because the related definition of bicoherence in this case can be regarded as a simple weighted average over bispectral phase [28,37], it allows for a particularly simple correction for small-sample and amplitude-related biases, which may otherwise degrade the filter by exaggerating the contribution of frequencies in which power is excessively skewed [34]. Specifically, bias in magnitude-weighted bicoherence, $\epsilon$, can be related to the inverse root of the "effective degrees of freedom" within the average, as

$$\epsilon = \sqrt{\frac{\sum w_i^2}{(\sum_i w_i)^2}} \tag{60}$$

where the "weights," are given by $w_i = |X_i(\omega_1)X_i(\omega_2)X_i(\omega_1+\omega_2)|$. In general, we have observed that the proposed algorithm is not overly sensitive to the choice of normalization, but the amplitude-weighted average with bias correction seems to perform well under a range of conditions. We have therefore defaulted to this form, in spite of the fact that it is perhaps the least principled of the three. The same arguments apply in the obvious way to polycoherence weighting with HOS of order K.

### 4.2. Partial delay filters

Consider the following filter obtained with record *j*:

$$r_{0j...j}(t) = \mathscr{F}^{-1}\left\{\int \ldots \int \hat{H}(\omega_1, \ldots, \omega_{K-1})X_j(\omega_2)\ldots X_j^*\right.$$

$$\left. \times \left(\sum_k^{K-1} \omega_k\right) d\omega_2 \ldots d\omega_{K-1}\right\}$$

$$\sim \hat{h} * (\hat{h} * x_j)^{K-1}(t) \tag{61}$$

The output of $h*x_j$ will be a symmetric linear-phase (i.e. zero-phase except for a time shift) function whose maximum is centered at the delay of *f* in the *j*th record along with the noise that survives filtering. To this term, Eq. (61) applies an exponent, causing the filter output to take on a spectrally broader and more impulse-like character, dominated, as K increases, by a sharp peak at the time of the maximum value within the record. Such a function is well approximated by an impulse shifted by $\hat{\tau}_j$ and weighted by $r_j^{K-1}(\hat{\tau}_j)$:

$$(h * x_j)^{K-1}(t) \sim \delta(t - \hat{\tau}_j) r_j^{K-1}(\hat{\tau}_j) \tag{62}$$

which becomes equality, almost surely, in the limit as *K* increases. We may therefore treat the filter as closely approximate to a weighted and shifted copy of the matched filter:

$$r_{0j...j}(t) \sim \hat{h}(t - \hat{\tau}_j) r_j^{K-1}(\hat{\tau}_j) \tag{63}$$

This filter can be used directly to estimate the delay between record *j* and another record, *i*:

$$r_{ij...j} = r_{0j...j} * x_i(t) \tag{64}$$

We will call it the *partial delay filter* for *j*.

An interesting alternative to delay estimation from (64) takes the estimate from the cross-correlation of partial filters:

$$r_{0jj...j} \star r_{0i...i} \sim (\hat{h} \star \hat{h})(t - \Delta\hat{\tau}_{ij})\left[r_i(\hat{\tau}_i)r_j(\hat{\tau}_j)\right]^{K-1} \tag{65}$$

This possibility is attractive because the result appears to contain less noise: the influence of noise in perturbing the peak location is effectively already "baked into the cake" within the respective partial filters, and the cross-correlation will at least give the impression of being relatively noise-free in comparison to (64). However, in considering HOS decompositions later we will find reasons why it makes sense to prefer (64).

### 4.3. Generalizations for complex and multivariate signals

The preceding arguments generalize to complex and multivariate signals with little difficulty, using properties described in sections 2.1.5 and 2.1.6. The models underlying these extensions assume that *m* component signals each contain a distinct feature with relative timing fixed across components in a background of correlated stationary noise. As described in Section 2.1.6, the application to a univariate complex signal, $y(t) \in \mathbb{C}^1$, is isomorphic to the bivariate case, $x(t) \in \mathbb{R}^2$, with implicit cross-spectra between negative and positive frequencies. In both cases the extension entails a summation over separate auto- and cross-spectra, which optimally accounts for the correlation structure of the noise both within and across the component signals. A more detailed development of these ideas is provided in Appendix A.

## 5. Application to non-deterministic HOS

So far we have mostly limited our attention to deterministic HOS. Applications to non-deterministic HOS are of interest for a wide range of problems, yet few, if any, prior treatments of signal recovery with HOS venture beyond the deterministic case, to our knowledge. For non-deterministic signals that result from the additive output of independent deterministic processes, cumulant HOS are likewise additive. Ignoring those slices in HOS space at which cumulant and moment HOS may diverge, we note that all of the crucial equations in the previous sections take on an additive form. Polycoherence weighting, under the Gaussian noise assumption, becomes:

$$H(\omega_1, \ldots, \omega_{K-1}) = \sum_{p=1}^{P} H_p(\omega_1) \ldots H_p(\omega_{K-1}) H_p^*\left(\sum_k \omega_k\right) \quad (66)$$

With quasi-cumulant windowing, the partial delay estimator in (61) decomposes accordingly

$$r_{0j\ldots j}(t) = \sum_{p=1}^{P} \hat{h}_p * \left[\left(\hat{h}_p * x_j\right)(t)^{K-1} - C\left[\hat{h}_p * x_j\right](t)\right]$$

$$\sim \sum_{p=1}^{P} \hat{h}_p(t - \hat{\tau}_{pj}) \alpha_{pj}^{K-1} \quad (67)$$

where $\alpha_{pj}$ is the peak value for process $p$ in record $j$. The term $C$ is introduced by the exclusion of regions where moment and cumulant HOS diverge and corresponds to the lower-order terms in the cumulant polynomial. It therefore involves a summation over terms of order less than $K - 1$. For the sake of economy, we will disregard these lower-order terms in the following discussion, leaving their effect implicit. Pairwise delay estimators thus become:

$$r_{ij\ldots j}(t) = r_{0j\ldots j} * x_i(t) \sim \sum_{p=1}^{P} (\hat{h}_p * x_i)(t) \alpha_{pj}^{K-1} \quad (68)$$

or the alternative form:

$$\left(r_{0i\ldots i} \star r_{0j\ldots j}\right)(t) \sim \sum_{p=1}^{P} \sum_{q=1}^{Q} (\hat{h}_p \star \hat{h}_q)(t - \Delta\hat{\tau}_{ij}^{pq})\left[\alpha_{pj}\alpha_{qi}\right]^{K-1} \quad (69)$$

Both clearly give a mixture of peaks related to the different generating processes, but as yet we have no way to sift peaks according to processes. In the spectral domain, Eq. (69) assumes the form of a tensor product, and so one might attempt to separate components through a variety of tensorial decompositions [62]. We will develop an alternative, computationally less expensive and, for most real-world settings, more practical approach, which we have found to work well with the bispectrum. This approach is particularly advantageous when handling large numbers of records and in real-time applications.

### 5.1. Iterated realignment

In this approach we start with the *average* partial delay filter:

$$g^{(0)}(t) = \frac{1}{L}\sum_{j=1}^{L} r_{0j\ldots j}(t) = \frac{1}{L}\sum_{j=1}^{L}\sum_{p=1}^{P} \hat{h}_p * (\hat{h}_p * x_j)^{K-1}(t)$$

$$\sim \frac{1}{L}\sum_{j=1}^{L}\sum_{p=1}^{P} \hat{h}_p(t - \hat{\tau}_{pj})\alpha_{pj}^{K-1}$$

$$\rightarrow \frac{1}{L}\sum_{p=1}^{P} \psi_p^{(0)} * \hat{h}_p(t) \mathrm{E}\left[\alpha_p^{K-1}\right] \quad (70)$$

where $\psi_p^{(0)}$ is the probability distribution of delays for process $p$, $\tau_p$, over records. In general, $\psi$ acts as a low-pass filter, attenuating energy as $1/\sqrt{L}$ above some frequency threshold reflecting the scale of variability in $\tau$. However, the sample distribution will not be smooth, but contain variations in the density of $\tau$'s, either random or related to modes in the underlying probability distribution, which will prove important later.

We will next obtain a provisional delay estimate from the time of the maximum value (for odd $K$, and maximum absolute value for even $K$) from each record:

$$\hat{\tau}_j^{(1)}(t) = \arg\max_t g^{(0)} * x_j(t) \quad (71)$$

and repeat the procedure after shifting records according to $\hat{\tau}^{(1)}$:

$$g^{(1)}(t) = \frac{1}{L}\sum_{j=1}^{L} r_{0j\ldots j}(t + \hat{\tau}_j^{(1)}) \quad (72)$$

The maximum in a given record will come from one of three sources: (1) a peak related to a feature emitted by process $p$, (2) a peak related to a process other than $p$, and (3) noise unrelated to any feature-emitting process. Peaks arising from process $p$ will tend to occur at delays aligning to random or distribution-related clusters within the sample distribution of $\tau_p$, whereas noise-related peaks and peaks unrelated to process $p$ will be as randomly distributed as before, relative to the sample distribution for $\tau_p$. To the extent that the shifted records align on modes in the sample distribution of $\tau_p$, the distribution of delays after realignment, $\psi_p^{(1)}$, will be sharpened, and its spectral threshold correspondingly extended. Because of this, the re-estimated filter $g^{(1)}$ becomes more closely matched to processes that capture the most maxima (assumed for argument to include $p$), and peak selection more strongly favors the feature associated with process $p$ at the next iteration, leading to an accelerating convergence towards $h_p$:

$$\hat{\tau}_j^{(m+1)}(t) = \hat{\tau}_j^{(m)} + \arg\max_t \left[g^{(m)} * x_j(t + \hat{\tau}_j^{(m)})\right] \quad (73)$$

$$g^{(m+1)}(t) = \frac{1}{L}\sum_{j=1}^{L} r_{0j\ldots j}(t + \hat{\tau}_j^{(m+1)}) \quad (74)$$

The likelihood of being among the favored processes at iteration $m$, depends on four factors: (1) the relative maximum signal-to-noise ratio, $\alpha_p$, (2) the variability of $\tau_p^{(m-1)}$, (3) the similarity of $h_p$ to $g^{(m)}$, and (4) the amount of energy in $h_p$ falling below the characteristic scale of $\psi_p^{(m)}$. If the timing distribution in all processes is uniform, then signal-to-noise ratio will tend to be the main initial determinant.

It is also evident that $g$ might converge on a filter that combines $h$ from different processes, which certainly must be the case in the limit of identical waveforms. More generally, waveforms will tend to be lumped together into common averages to the extent that the matched filter for a given process, $h_p$ produces a peak in the output, $h_p * f_q$, which stands out from noise reliably enough to produce the maximum value in a subset of records. The ability to distinguish among features therefore depends on the similarity between the respective waveforms, their common resemblance to $g$, and the presence, magnitude and spectral properties of noise. This fact is the main reason to favor (64) as a delay estimator over (65): the noise within individual records aids in discriminating among features. Stochastic resonance therefore plays a role in feature separation.

An example of filter estimation and its convergence is shown in Fig. 1.

## 6. Signal reconstruction

Once the foregoing algorithm has converged on a suitable filter, a component signal may be reconstructed by (1) recovering the waveform using estimated occurrence times (2) applying the filter to any relevant records, (3) thresholding the resulting output, (4) convolving the recovered waveform with the thresholded signal, and (5) scaling the resulting component signal to minimize squared error, as explained next.

### 6.1. Waveform recovery

The underlying waveform may be recovered by averaging over records aligned to peaks in the detection filter. One strategy is to obtain a waveform as a simple average over all records aligned to the maximum value in the output of the filter:

$$\hat{f}_j(t) = \frac{1}{L}\sum_{i=1}^{L} x_i(t + \hat{\tau}_{ij}) \tag{75}$$

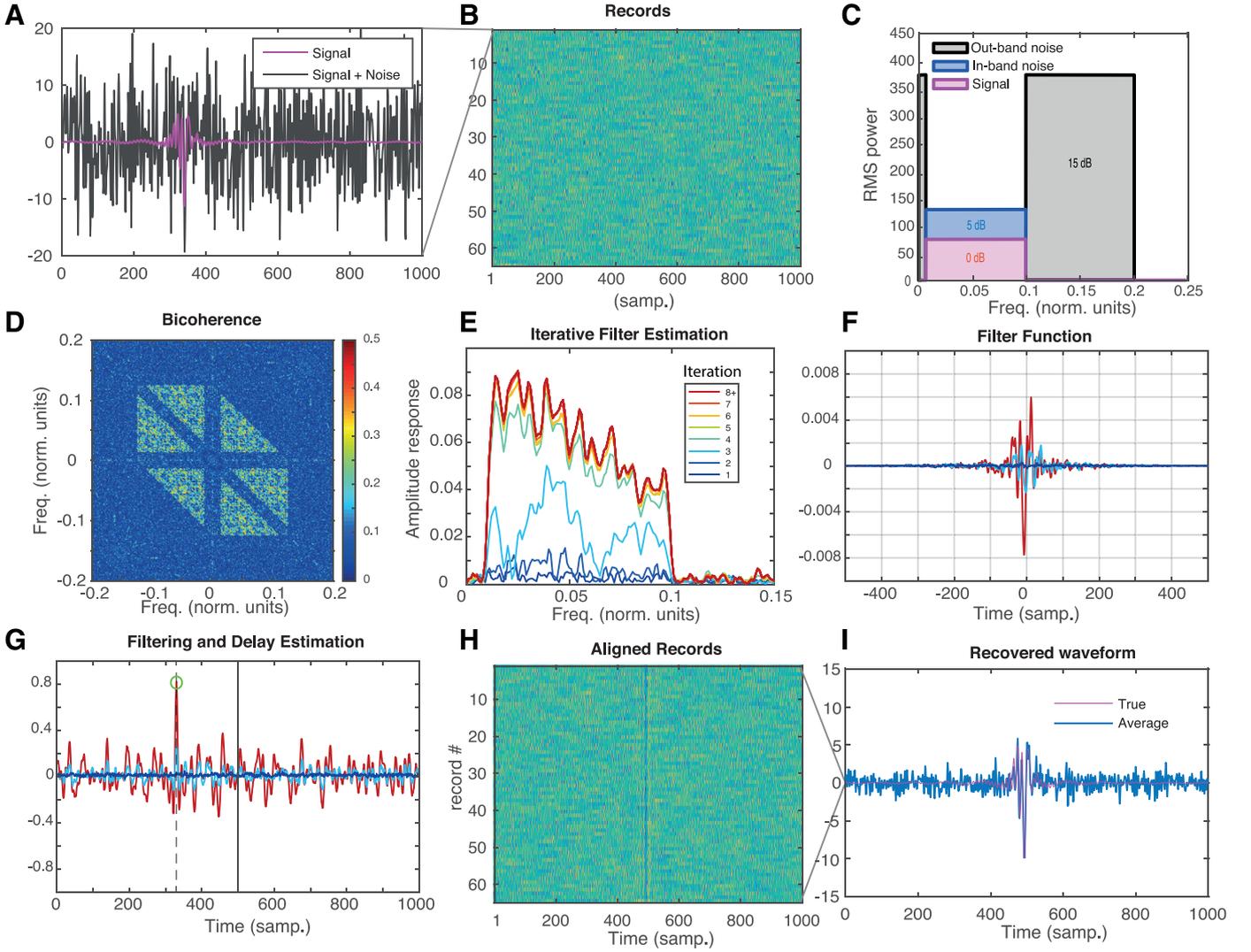

**Fig. 1.** Recovery of test signals with bispectral delay estimation. Test records were arrays of randomly delayed instances of the same test signal, exemplified in panel **A** (*magenta*), embedded in independent spectrally colored noise (**A**, *black*: test signal with 9.5 dB noise). **B**: example array with 64 records. The test signal was designed with a flat power spectrum within a specified frequency range (example case, **C** *magenta*), while noise amplitude was varied separately within the signal bandwidth (in-band noise, **C**, *blue*) and outside (out-band noise, **C**, *black*). The delay filter was obtained from cross-bicoherence averaged over all pairs of records, weighted by the complex-conjugate of (auto-)bicoherence averaged over records (**D**). The procedure was repeated 15 times; at each iteration, average cross-bicoherence was re-estimated after aligning records according to the delay estimated from the previous iteration, and the filter updated. Convergence is rapid, typically requiring fewer than 10 iterations, as reflected in the amplitude response of the filter (**E**), the filter function (**F**) and the efficacy of the filter in detecting the signal (**G**; filter applied to the first record shown in **A**). **I**: Aligned records. The signal was recovered from the average over aligned records (**H**). (For interpretation of the references to color in this figure legend, the reader is referred to the web version of this article.)

This approach weights all records uniformly in the average, and is suitable if one assumes the feature to be generally prevalent across records. We have already obtained the detection filter through a weighted averaging over records in (70); therefore an unequally weighted average might be recovered from the delay filter itself by applying the inverse of the estimated noise amplitude spectrum to the complex conjugate of the filter spectrum. Finally, a waveform might be recovered by deconvolving the filtered signal from the observed signal, treating the former as a proxy for a known input [65].

### 6.2. Signal estimation

We have in hand, so far, a waveform estimate and the output of the detection filter, and we wish to reconstruct the signal according to our model in (1). This might be done by directly convolving the output of the detection filter with the estimated waveform [15,65], which effectively returns the original signal filtered according to the signal to noise ratio over frequency. But because we have information that localizes the signal in time as well as frequency, we may do better by windowing in both domains. We therefore propose to convolve the estimated waveform with a windowed copy of the signal.

$$\hat{y}_{ij}(t) = \int \hat{f}_j(t-s) w[g_j * x_i](s)\, ds \tag{76}$$

To obtain the proper expectation according to the signal model in (1), we must, ideally, convolve $f$ with the expected distribution of occurrence times, multiplied by the expected scaling. We therefore seek a windowing that approximates this distribution: $w[g_j * x_i](t) \approx \alpha_{ij} \psi_{ij}(t)$.

Properties of this distribution may be inferred directly from the output of the detection filter. The peak magnitude clearly provides information on the likelihood that the signal is present. The probability of a feature at $t$ is given by a prior probability and the likelihood of the filter output, tending naturally to be a sigmoidal function of output, or absolute value in the case of even orders. The sigmoid is governed by a threshold parameter, which adjusts to the prior probability of an emission, and a slope parameter. We may also infer something about the local ambiguity in the timing of the feature on the basis of the second derivative at the peak; in fact, Eq. (43) implies that the shape of the peak already approximates that of the local distribution of peak times, thus the filter output itself may serve the desired purpose without further modification. A sensible form for $w$ therefore applies the first term as a sigmoidal nonlinearity, $s_\theta$, to the second term, which is the output itself:

$$w[g_j * x_i](t) = s_\theta(g_j * x_i(t))(g_j * x_i)(t) \tag{77}$$

By this argument, we may justify windowing the output according to magnitude. As the slope parameter of the sigmoid is of secondary importance, it will be ignored for the present, and we will take $s_\theta$ to be a hard threshold function. How to set the threshold is considered next.

### 6.3. Threshold determination

Several different methods might be applied towards selecting a threshold, including an explicit modeling of the likelihood considered earlier. But in the context of HOS decomposition, the following simple criterion has a clear motivation: choose a threshold such that the $K$th-order scalar cumulant computed over all subthreshold samples vanishes, or otherwise attains some statistically-motivated value. Such a threshold addresses the null-hypothesis that the data contain only Gaussian noise, and can be understood roughly as orthogonalizing the residual signal in the HOS space, producing an outcome analogous to the mean-square error criterion in standard regression.

Applying the idea to the bispectral case, denote $\langle \ldots \rangle_{\lceil \theta \rceil}$ as an average excluding values above a threshold value, $\theta$, and compute skewness as

$$\gamma_{\lceil \theta \rceil} = \frac{\langle r^3(t) \rangle_{\lceil \theta \rceil} - 3 \langle r^2(t) \rangle_{\lceil \theta \rceil} \langle r(t) \rangle_{\lceil \theta \rceil} + 2 \langle r(t) \rangle_{\lceil \theta \rceil}^3}{\left( \langle r^2(t) \rangle_{\lceil \theta \rceil} - \langle r(t) \rangle_{\lceil \theta \rceil}^2 \right)^{3/2}} \tag{78}$$

Making use of the large-sample standard deviation of skewness, we may attempt to limit the false positive rate to some predetermined value, FP, by setting $\theta > 0$ such that

$$\gamma_{\lceil \theta \rceil} < \Phi^{-1}(1 - \text{FP})\sqrt{6/T} \tag{79}$$

where $T$ is the record duration in samples and $\Phi^{-1}$ is the inverse standard normal cumulative distribution function. For example, to limit the probability of a false detection in the absence of a signal to approximately 5%, set

$$\gamma_{\lceil \theta \rceil} < 1.64\sqrt{6/T} \tag{80}$$

### 6.4. Decomposition

Decomposition proceeds by way of the "deflation" technique [15,22]: a component signal is estimated and subtracted from the original signal, and the entire procedure repeated with the residual. The number of components may be determined in a data-driven manner by continuing the decomposition until the zero-lag moment of the estimate on the residual signal falls below some statistically motivated threshold.

## 7. Relationship to MED

It was claimed earlier that the HOS delay estimators relate to moment-maximizing filters of the type often applied to the problem of blind deconvolution [10,11,24,51,52,67]. Minimum entropy deconvolution (MED) seeks a filter which maximizes a moment normalized for scale invariance; most commonly kurtosis [10,24]:

$$\gamma_K = \frac{\frac{1}{L} \int \sum_{j=1}^{L} (h * x_j)^K(t)\, dt}{\left[ \frac{1}{L} \int \sum_{j=1}^{L} (h * x_j)^2(t)\, dt \right]^{K/2}} \tag{81}$$

The samples indexed by $j$ might be shorter possibly overlapping windowed intervals of fixed duration from some longer record, with a duration that will determine the filter order of $h$. For the present purpose, we will formulate an equivalent Lagrangian objective function, leaving the normalization implicit within an energy constraint.

$$\Lambda = \int \frac{1}{L} \sum_{j=1}^{L} (h * x_j)^K(t)\, dt + \lambda \left( 1 - \int \frac{1}{L} \sum_{j=1}^{L} (h * x_j)^2(t)\, dt \right) \tag{82}$$

where $\lambda$ is here a Lagrange multiplier.

Expanding (82) in the frequency domain:

$$\Lambda = \int \ldots \int \hat{M}^K(\omega_1, \ldots, \omega_{K-1}) H(\omega_1) \ldots H(\omega_{K-1})$$
$$\times H^* \left( \sum_{k=1}^{K-1} \omega_k \right) d\omega_1 \ldots d\omega_{K-1} + \lambda \left( 1 - \int |H(\omega)|^2 \hat{M}^2(\omega) d\omega \right) \tag{83}$$

where $\hat{M}^K$ is the $K$th-order spectral estimate

$$\hat{M}^K = \frac{1}{L} \sum_{j=1}^{L} X_j(\omega_1) \ldots X_j(\omega_{K-1}) X_j^* \left( \sum_{k=1}^{K-1} \omega_k \right) \tag{84}$$

At extrema in $\Lambda$, the derivatives vanish, $\left. \frac{\partial \Lambda}{\partial |H(\eta)|} \right|_{H=\hat{H}} = 0$ giving

$$\hat{H}^*(\eta) \propto \frac{1}{\hat{M}^2(\eta)} \int \ldots \int \hat{M}^K(\eta, \omega_2, \ldots) \hat{H}(\omega_2) \ldots$$
$$\times \hat{H}^* \left( \eta + \sum_{k=1}^{K-1} \omega_k \right) d\omega_2 \ldots d\omega_{K-1} \tag{85}$$

Or in its more economical time-domain expression:

$$\hat{h}(t) \propto u^{(-1)} * \frac{1}{L} \sum_{j=1}^{L} x_j * (\hat{h} * x_j)^{K-1}(-t) \tag{86}$$

where $u^{(-1)}$ is meant to indicate the whitening filter that results from $1/\hat{M}^2$ in (85).

There are a few things to unpack from Eq. (86). We have already observed that if $x$ contains some deterministic feature, $f$, in stationary Gaussian noise, the phase response of any moment-maximizing $h$ should be matched to the phase spectrum of $f$ (Lemma 1). We have also seen in Eq. (62) that the exponent in (86) results in an impulse-like function at the delay of $f$ within the record. We may therefore treat the net effect of Eq. (86) as closely approximate to a weighted summation over records aligned to peaks in the filter output. If multiple peaks of sufficiently similar amplitude appear in a given record, then the summation includes multiple correspondingly shifted copies of the record, although the tendency as $K$ increases will always be for a single peak to dominate.

## 7.1. Advantages of HOSD over MED

Comparing HOSD and MED, some advantages of HOS-based techniques over direct maximization of (82) become apparent. The main ones are:

### 7.1.1. No need for numerical optimization

Gradient ascent of (82) typically begins with some arbitrary initial value of $h$, which in most cases is likely to be far from any optimum. But we have already found enough information in HOS to recover $h$ from a deterministic spectrum, at least implicitly within the bicoherence-weighted partial delay estimator of (61). Once the relative delays are in hand, explicit recovery of both the filter and the underlying waveform becomes a matter of averaging over aligned records. All of this we have gained without recourse to any kind of gradient-based numerical optimization over $h$.

### 7.1.2. Improved outlier sensitivity

Because the sum over shifted records in Eq. (86) weights according to $r_j^{K-1}(\hat{\tau}_j)$, the record with the greatest maximum comes to dominate in the summation as $K$ increases. More generally, the filter estimate will be heavily weighted towards any outliers, often impractically so. One consequence of this is a tendency for MED to converge on degenerate solutions: it is always possible to design a filter matched to a particular segment of the input, which produces one large impulse with high kurtosis. The approach developed here adds more options for mitigating this problem.

### 7.1.3. Greater flexibility

With MED, weighting of HOS is limited to the magnitude of the deterministic HOS of $h$. We have just obtained delay estimators by designing windows directly within the higher-spectral domain. For example, bicoherence weighting described in Section 4.1, does not necessarily correspond to any deterministic spectrum. This allows for considerably greater flexibility in filter design, which, in the case of bicoherence weighting, improves optimization under forms of non-Gaussian WSS noise.

### 7.1.4. Separation of additive processes

Strategies for separating additive processes become more obvious when approaching the problem through HOS.

## 8. Methods: Algorithms

### 8.1. Implementation of HOSD

The following section implements this framework in an algorithm which recovers the delays between $L$ discretely sampled records, which contain $f$ at mutually independent lags in a background of independent Gaussian noise with identical power spectra.

#### 8.1.1. Windowing

The first step is to form the array of records. If these are obtained as shorter intervals of a longer record, then a record duration must be chosen based on the time scale of the signal of interest and computational considerations. It will also make sense to apply a window to the record to suppress spectral leakage in the estimators and related artifacts. In the following it is assumed that a suitable window has already been applied to each record, $x_j$.

#### 8.1.2. Bispectral weighting

In this setting, a consistent estimator of the auto-bispectrum is obtained by averaging over channels, which will be done under the assumption that noise is identically distributed across records. Likewise, an estimate of the denominator is obtained by averaging the respective terms, which may be plugged in to Eq (55), giving the bicoherence filter:

$$\hat{H}[\omega_1, \omega_2] = \frac{\hat{B}^*[\omega_1, \omega_2]}{D[\omega_1, \omega_2]} \quad (87)$$

where $B$ is a direct bispectral estimate:

$$\hat{B}[\omega_1, \omega_2] = \frac{1}{L}\sum_{j=1}^{L} X_j[\omega_1] X_j[\omega_2] X_j^*[\omega_1 + \omega_2] \quad (88)$$

with $X_k[\omega]$ the fast Fourier transform (FFT) of $x[t]$ and $D$ normalizing either in accordance with the variant of the common definition of bicoherence in Eq. (58):

$$D_{BC}[\omega_1, \omega_2] = \frac{1}{L^2}\sum_{i=1}^{L} |X_i[\omega_1]|^2 \sum_{j=1}^{L} \left|X_j[\omega_2] X_j^*[\omega_1 + \omega_2]\right|^2 \quad (89)$$

or with the magnitude-weighted definition:

$$D_{MW}[\omega_1, \omega_2] = \left[\frac{1}{L}\sum_{j=1}^{L} \left|X_j[\omega_1] X_j[\omega_2] X_j^*[\omega_1 + \omega_2]\right|\right]^2 \quad (90)$$

In the case of magnitude weighting, we may correct for bias as described in (60) [34,37]

$$\hat{H}_\epsilon[\omega_1, \omega_2] = \frac{\hat{B}^*}{\sqrt{D_{MW}}}\left[\frac{1}{\sqrt{D_{MW}}} - \frac{\epsilon}{|B|}\right] \quad (91)$$

The window, $H$, may be modified in other ways according to any relevant considerations. For example, computational efficiency might be served by limiting the estimator to specific signal-relevant bandwidths within the HOS domain, or properties of the estimator improved by accounting for prior information through the weighting.

#### 8.1.3. Delay estimation with averaged cross-bispectra

Eq. (87) gives us a suitably consistent estimate of the bispectral filter needed for partial delay filter estimation. Following Eq. (74), the average partial delay filter, $g$, is found by iterating

$$G^{(m+1)} = \frac{1}{L}\sum_{j=1}^{L} e^{i\omega_1 \hat{\tau}_j^{(m)}} \sum_{\omega_2=-W}^{W} X_j[\omega_2] X_j^*[\omega_1 + \omega_2] \hat{H}[\omega_1, \omega_2] \quad (92)$$

where $\hat{\tau}_j^{(m)}$ is the estimated lag of the delay in the $j$th record after the $m$th iteration, obtained as in (73). Even when the distribution of $\tau$'s is initially uniform, we have found in practice that the tendency to cluster around random modes early in the iteration leads to rapid convergence, as exemplified in Fig. 1. When the algorithm successfully converges, noise is suppressed as

$$G[\omega_1] \to \sum_{\omega_2=-W}^{W} F[\omega_2] F^*[\omega_1 + \omega_2] H[\omega_1, \omega_2] + O\left(\frac{1}{\sqrt{L}}\right) \quad (93)$$

Which converges to a filter that is approximately the optimal linear filter. For the Gaussian case, we have:

$$g(t) \to h * (h \star h)^2(t) \quad (94)$$

Again, because $(h\star h)^2$ is an impulse-like function, performance of $g$ typically approaches that of $h$, and we may regard $g$ and $h$ as functionally equivalent in settings that involve spectrally broad features of the type represented in the bispectrum.

#### 8.1.4. Feature recovery

Following iterated delay estimation, the feature, $F$, is recovered from the delay-compensated average:

$$\hat{F}(\omega) = \frac{1}{L}\sum_{j=1}^{L} e^{i\omega \hat{\tau}_j^{(m)}} X_j(\omega) \quad (95)$$

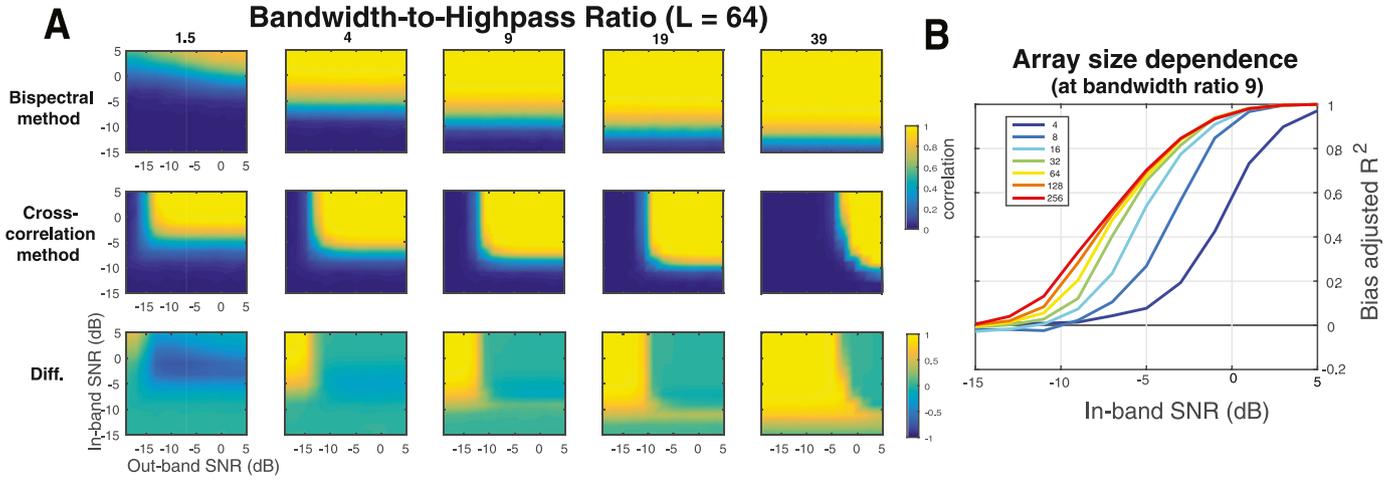

**Fig. 2.** Comparison of bispectral and 2nd-order delay estimators and their dependence on noise and signal bandwidth. Performance was evaluated over 400 randomly generated signal arrays at multiple combination of in- and out-band noise level, as described in Fig. 1. The test metric for the comparison was the correlation between true and estimated delays. **A**: Performance of the bispectral algorithm diminished only with increasing in-band noise, but was not affected by out-band noise (**A**, first row). In contrast, the effectiveness of the cross-spectral algorithm diminished both with increasing in-band and out-band noise (**A**, second row). The relative advantage of the bispectral method grew with increasing bandwidth of the test-signal (**A**, third row). **B**: Robustness to in-band noise also improved with increasing array size (shown for bandwidth ratio 9).

#### 8.1.5. Real-time and running-average estimation

Because we have derived the filter as a simple average over records, the general scheme adapts easily to real-time applications. In the simplest form, real-time applications may apply a continuously updated estimate of the filter towards detecting the signal within a buffer containing an incoming data stream. Specifically, data written to a buffer at some interval, $\Delta T_{m+1}$, are filtered with $g^{(m)}$ and shifted according to the delay estimate, $\hat{\tau}_m$, obtained as before. The convergence to a matched filter here arises from the serial application to incoming records rather than by iterating through all records. With each record $H$ may be updated as a running averages according to

$$H^{(m+1)} = \frac{B^{(m+1)}}{D^{(m+1)}} \tag{96}$$

where

$$B^{(m+1)} = (1 - \lambda[\delta_m])B^{(m)} + \lambda[\delta_m](X_m[\omega_1]X_m[\omega_2]X_m^*[\omega_1+\omega_2]) \tag{97}$$

and, using magnitude-weighted normalization

$$\sqrt{D^{(m+1)}} = (1 - \lambda[\delta_m])\sqrt{D^{(m)}} + \lambda[\delta_m]|X_m[\omega_1]X_m[\omega_2]X_m^*[\omega_1+\omega_2]| \tag{98}$$

The delay-estimating filter, $g$, is similarly updated as

$$G^{(m+1)} = (1 - \alpha[\delta_m])G^{(m)} + \alpha[\delta_m]P_m \tag{99}$$

with

$$P_m[\omega_1] = e^{i\omega_1\hat{\tau}_m} \sum_{\omega_2=-W}^{W} X_m[\omega_2]X_m^*[\omega_1+\omega_2]H^{(m+1)}[\omega_1,\omega_2] \tag{100}$$

where $\lambda[\delta_m]$ and $\alpha[\delta_m]$ are learning rates within [0,1], which may be fixed or set to depend on the detection parameter $\delta_m$. For example, a hard threshold might be set on the maximum in $r_m$:

$$\delta_m = \begin{cases} 1, & \text{if } \max r_m > \theta_m \\ 0, & \text{otherwise} \end{cases} \tag{101}$$

with $\lambda[0] = \alpha[0] = 0$.

#### 8.1.6. Decomposing multi-component signals and signals in non-Gaussian noise

Real-world applications must often confront noise that is not Gaussian, whose non-vanishing HOS may mask that of the signal. To address non-Gaussian noise as well as interference from competing signals, a decomposition may be implemented by repeatedly applying filter estimation, thresholding and signal reconstruction, as described in Section 6. The scheme might be implemented through a cascade of adaptive filters, each receiving the residual of the output from the preceding filter.

#### 8.1.7. Computational efficiency

For 3$^{rd}$-order HOS, the summation over $\omega_2$ in (92) is quadratic in $W$, but in the case of an ensemble of records it needs to be carried out only once at the outset and might be done intermittently in the case of real-time applications. Each subsequent step of the algorithm is linear in $W$. The efficiency of the algorithm therefore compares favorably with cross-spectral methods of delay estimation, which scale quadratically in the number of records at each iteration. Efficiency of the bispectral algorithm should surpass that of pairwise cross-spectral delay estimation whenever $W^2 < L^2$, as is common in applications to pattern recovery from long-duration records. More generally, computational complexity grows as $W^{K-1}$, reflecting the number of frequency dimensions, so that complexity increases at a super-quadratic rate for orders greater than 3.

### 8.2. ECG wavelet filter

Blind bispectral feature recovery was benchmarked against a representative wavelet filter of the type commonly used in heartbeat detection and denoising [16] in ECG. This was carried out by decomposing the noise-corrupted test signal with the "Sym4" discrete wavelet transform, then applying a hard universal threshold [25] to coefficients in scales 3–5 and discarding all other coefficients. QRS times were obtained from peaks in the denoised signal.

### 8.3. Alternative MED algorithms

Heartbeat detection was used also in comparing two implementations of kurtosis-maximizing MED to the filter estimation step of the HOS decomposition (HOSD) algorithm. The first is described in [45], and implemented in a Matlab program (Med2d.m)

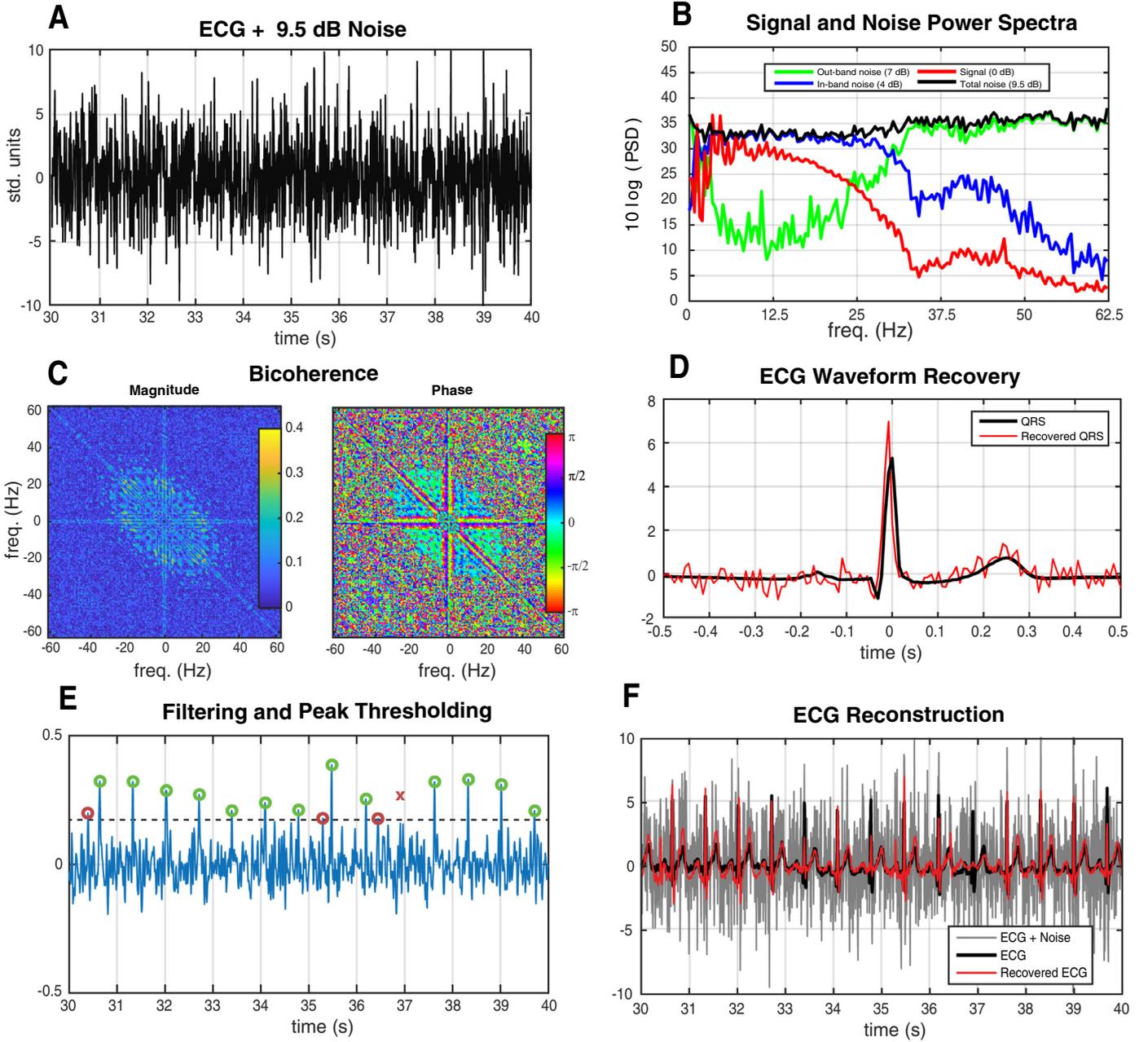

**Fig. 3.** Blind recovery of electrocardiogram (ECG) under large-amplitude noise. Eighteen one-minute samples of normal ECG, obtained from the MIT-BIH Normal Sinus Rhythm Database (nsrdb), served as test signals to which varying levels of in- and out-band noise were added (example in **A** with 7 dB out-band and 4 dB in-band noise added to record nsrdb/16483; spectral distributions of signal and noise components shown in **B**). The continuous ECG was divided into overlapping windows spanning approximately 3 R-R intervals and weighted by a Hann window, after which bispectral delay estimation was applied as described previously. Auto-bicoherence (**C**) recovers spectral characteristics of the signal used by the delay filter. Following realignment of all intervals, the ECG waveform was recovered from the delay-compensated average (**D**). The delay filter was then applied to the continuous signal and a peak threshold (**E**, dashed line) applied to identify occurrence times of the waveform (**E**, green circles indicate correct detections; red circles, false positives; and red x, missed detections). A reconstructed ECG (**F**) was obtained by convolving the recovered waveform with the peak-thresholded signal.

[43]. The second, described in [15], was likewise implemented in Matlab (Deflation.m) [13]. The Deflation algorithm also executes a decomposition of the higher-order moment through deflation; but it requires at least as many sensors as sources, hence was in this case only applied on a single channel assuming a single source. Because both algorithms maximize kurtosis, they were compared to the 4th-order variant of HOS decomposition, using the trispectrum (HOSD4), in addition to the bispectral version (HOSD3).

### 8.4. Second-order delay estimation

To compare the performance of second-order estimators to the bispectral method, pairwise delays were estimated for all records from the maximum value in respective cross-correlations. An array was then constructed encoding lags for each pair of records as a phase value:

$$\Phi = \left[\exp\left(2\pi i \frac{\Delta \tau_{jk}}{N}\right)\right]_{K \times K} \tag{102}$$

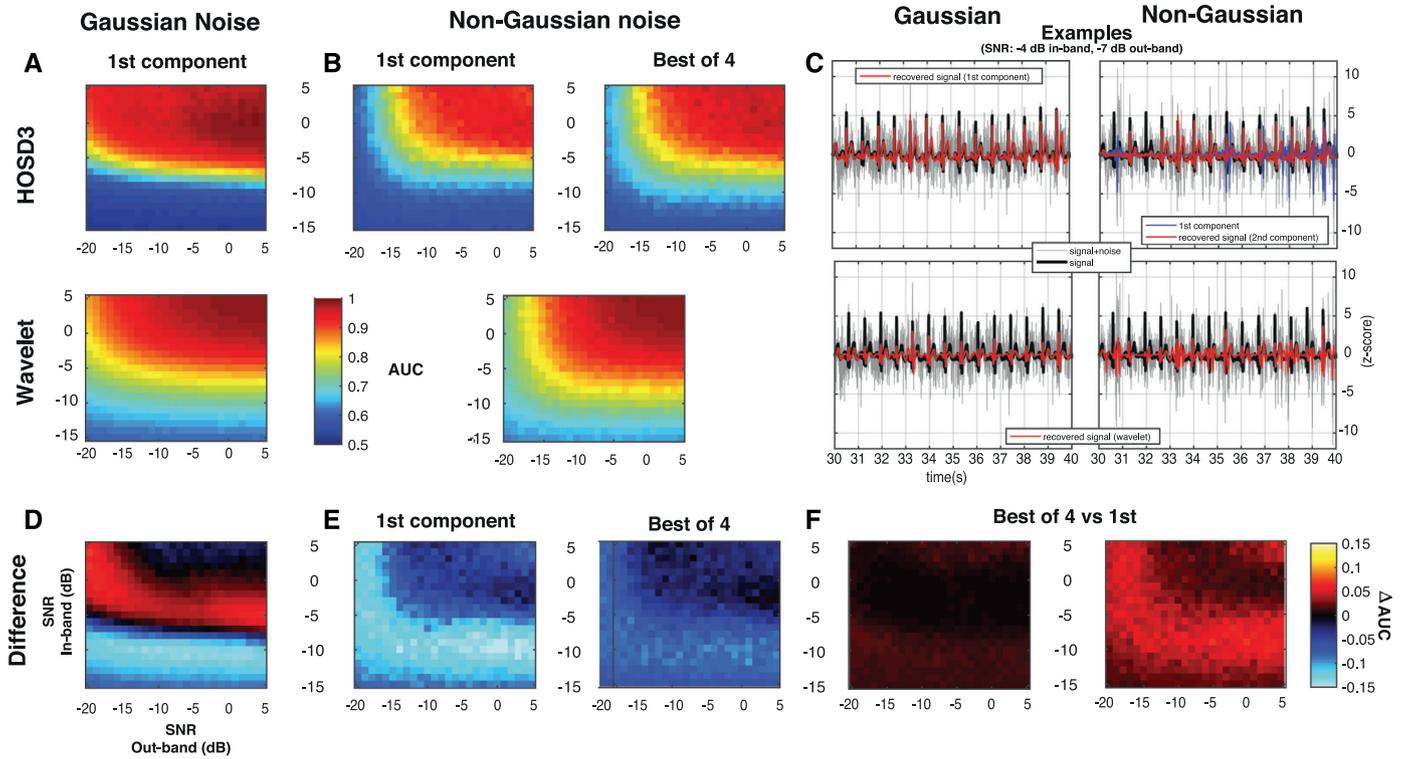

**Fig. 4.** Performance of the 3rd order (bispectral) decomposition (HOSD3) (**A,B,C**, top row) in heart beat detection, compared to a representative, non-blind (i.e. ECG-tailored) wavelet filter (**A,B,C**, bottom row), with additive Gaussian noise (**A**) and additive non-Gaussian noise (**B**). Note that the bispectral decomposition has not been expressly optimized for QRS detection, in contrast to the wavelet filter. Performance was measured at varying levels of spectrally matched ("in-band") and spectrally complementary ("out-band") noise using area-under-the-curve (AUC) computed over 5 peak-detection threshold levels. 20 noise realizations were added to each of 16 2-minute data samples obtained from the MIT-BIH normal sinus rhythm database; the average performance over all data sets and realizations is displayed in **A** and **B**. Non-Gaussian noise was generated by filtering squared Gaussian white noise with a non-minimum-phase filter whose amplitude response matched in- and out-band spectra and whose randomized phase response was smoothed to preserve the approximate time envelope of the respective zero-phase filters. Example cases are shown for Gaussian (**C**, left) and non-Gaussian (**C**, right) noise. For the non-Gaussian example, the ECG signal was recovered in the 2nd bispectral component (**C**, top right; red), while the first converged on features in the noise (**C**, top right; blue). Performance of both algorithms degrades beyond about −7 dB in-band SNR; the bispectral method exhibits a sharper transition, leading it to outperform wavelet QRS detection above the in-band threshold and underperform beyond the threshold (**D**). With non-Gaussian noise, the wavelet algorithm outperforms the first component (**E**, left); however, the best component of the first 4 performs comparably to the wavelet algorithm (**E**, right). Randomization of the filter phase introduces variability in the SNR performance threshold for both wavelet and bispectral methods, depending on the similarity between the filter waveform and the QRS complex, which leads to greater similarity in the average performance of the two methods under non-Gaussian noise. Difference between the first and best component performance under Gaussian noise (**F, left**) is negligible, in contrast to the difference for non-Gaussian noise (**F, right**), demonstrating that the improvement is not merely a result of bias from the selection of the best component.

where $N$ is the record length. Given that

$$E[\Phi] = \left[\exp\left(2\pi i \frac{\tau_j}{N}\right)\right]_{K \times 1} \left[\exp\left(-2\pi i \frac{\tau_k}{N}\right)\right]_{K \times 1}^T \quad (103)$$

the delays within individual records may be estimated from $\Phi$ according to the phases of the respective loadings onto the first component returned by a singular-value decomposition (SVD). Delays recovered in this way are unique up to an arbitrary constant time shift in all records, which does not affect the circular correlations used in comparing the algorithms.

*8.4.1. Woody's algorithm*

A variation of the cross-correlation technique, described by Woody [68], also served as a comparison. In general we observed that the aforementioned SVD technique gave qualitatively similar results to Woody's algorithm, with Woody's algorithm slightly more robust to in-band noise.

## 9. Methods: test signals

The test signals used in Fig. 1 were generated by applying a square FIR bandpass filter with a passband of 0.01 to 0.1, in units of normalized to sampling frequency, to windowed Gaussian white noise, where a Gaussian window with a standard width of 20 samples was applied in the time domain. The test signal was then normalized to have zero mean and unit variance. The result of this procedure is a nearly transient waveform with the desired properties, as illustrated in Fig. 1A. 64 records were created by embedding the test signal at mutually random and independent delays within Gaussian noise composed of two parts: "in-band" noise was constructed with a power spectrum matching that of the test signal by filtering white noise with unit variance, while "out-band" noise was obtained by subtracting the in-band noise from the same white noise used to generate the former, then low-pass filtering below 0.1 normalized units. Delays were implemented by circularly shifting each record by an amount drawn from a uniform distribution over the duration of the record. The demonstration was carried out in the circular domain for illustrative purposes, as this allows the respective finitely sampled bands to have ideal characteristics and simplifies the comparison of algorithms across conditions, which may be done with circular statistics.

Test signals used in Fig. 2 were generated in the same way at different combinations of the target signal bandwidth, in-band noise and out-band noise. The transient target signals were generated with bandwidths of 1.5, 4, 9, 19 and 39, as a ratio of the high-pass frequency of 0.01. Out-band signal-to-noise ratio varied from 5 dB to −15 dB, while in-band noise varied from 5 dB to −20 dB, where decibel values refer to the ratio between total signal energy and total noise within each band, separately. For each combination of bandwidth, in- and out-band noise levels, 400 such test-signal arrays were generated, and the results summarizing the average performance of bispectral delay estimator are shown in Fig. 2. Test signal generation and all analyses were programmed in Matlab using scripts and algorithms developed by the first author.

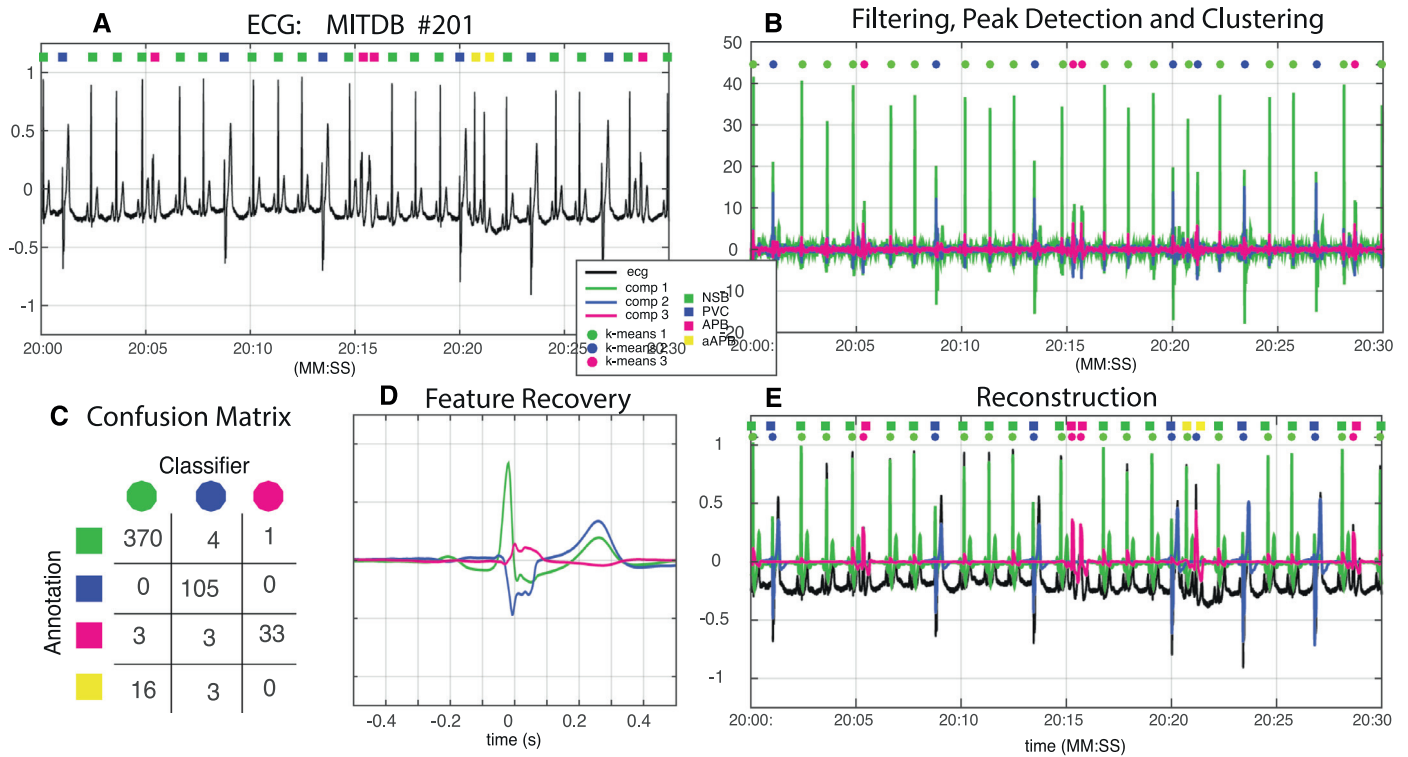

**Fig. 5.** Blind separation of normal from abnormal heart beats through the bispectral decomposition. The bispectral decomposition was applied to a 10 minute sample of ECG from the MIT-BIH Arrhythmia Database [27] (**A**; 00:15:00 to 00:25:00 of record #201 in 'mitdb'). Clinical annotation is indicated with colored squares: green, normal sinus beats (NSB); blue, premature ventricular contractions (PVC); magenta, atrial premature beats (APB); and yellow, aberrant atrial premature beats (aAPB). **B**: Filtered ECG for the first 4 components of the bispectral decomposition. Peaks in the filter output were classified with K-means clustering on component rms values within 70 ms. Circles indicate classifier output for each peak with colors matching the most closely associated clinical annotation for both the classifier output and the component activation. Energy in the first component is associated mainly with NSB's; the 2nd with PVC's; and the 3rd with APB's. **C**: Confusion matrix comparing the result of blind classification to clinical annotation, showing a close association between the classifier performance and clinical annotation for all beat types except aberrant premature atrial beats (aAPB), which did not cluster separately from NSB. Classifier performance did not improve with more than 3 clusters or the inclusion of more than 4 bispectral components. **D**: Recovered waveforms for the first three components. **E**: Reconstruction of component signals and side-by-side comparison of the clinical annotation (squares) and classifier output (circles). (For interpretation of the references to color in this figure legend, the reader is referred to the web version of this article.)

#### 9.0.2. ECG test signal

The performance of the algorithm in recovering a physiological signal from varying levels of noise was tested by adding random noise to 2 minute segments ECG obtained from the MIT-BIH normal sinus rhythm database [27]. Data were retrieved for 16 subjects, starting at sample number 10,000 (00:01:18.125 to 00:03:18.125), normalized to unit variance and divided into records of approximately 3-beat duration. Gaussian and Non-Gaussian noise were generated at multiple in- and out-band levels. The in-band noise spectrum was defined as

$$\sigma_{in} = \frac{|Y(\omega)|^2}{|Y(\omega)|^2 + 1}$$

where $Y$ is the spectrum of the input signal before adding noise, and

$$\sigma_{out} = \frac{1}{|Y(\omega)|^2 + 1}$$

To each array of records and at each combination of in- and out-band noise, 20 random noise realizations were added. Non-Gaussian noise was generated by applying a filter with amplitude response given as above to squared Gaussian white noise, whose distribution, $\chi^2(1)$, is highly skewed. To make the phase of resulting noise HOS independent of the signal HOS, the phase response of the filter was randomized, after which the filter function was windowed in the time domain so that the dispersion of energy in time resembled that of the zero-phase filter. Example data are displayed in Fig. 3, while performance averaged across subjects and realizations, in comparison to a representative wavelet filter, is displayed in Fig. 4.

## 10. Experimental results

The method is first demonstrated with arrays of randomly delayed test-signals embedded in Gaussian noise (Fig. 1), in comparison to alternative cross-spectral techniques (Fig. 2). As summarized in Fig. 2, the bispectral algorithm dramatically outperforms delay estimation based on pairwise cross-spectra when the signal power-spectrum cannot be recovered from noise. To illustrate this point, noise was varied separately in two bands: the first, "in-band noise," was designed to have a power spectrum which matched that of the signal and the second, "out-band noise," had the complementary spectrum. Because the total signal-plus-noise power happens to match the signal-only power for in-band noise, the bispectral algorithm should not greatly outperform the cross-spectral approach for a signal corrupted by in-band noise alone. A different outcome is expected for out-band noise because the cross-spectral approach fails to correctly reconstruct the signal spectrum. Fig. 2A shows the performance of the respective algorithms using the circular correlation between the true delays and estimated delays. As expected, the cross-spectral delay estimator failed at high out-band noise, with performance diminishing between $-5$ and $-10$ dB. In contrast, the bispectral algorithm shows virtually no decrement with increasing out-band noise, implying that the algorithm has no trouble distinguishing the signal-containing band even when total noise energy in the out-band exceeds that of the signal by a hundred fold. In this example, the separation of noise energy into in-band and out-band illustrates the how the algorithm identifies regions of the spectrum where signal-to-noise ratio is greatest.

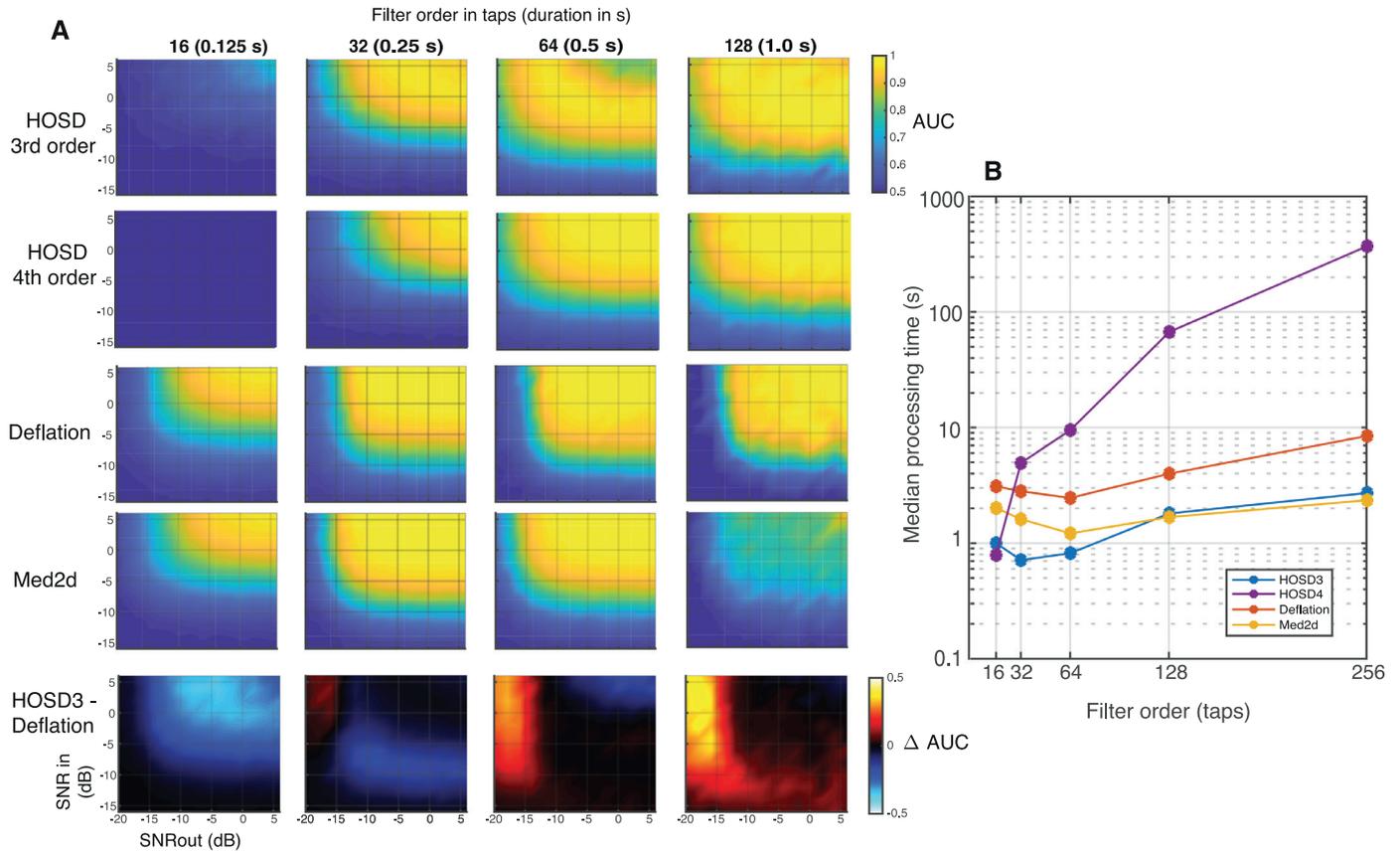

**Fig. 6.** Comparison of filter estimation techniques in heartbeat detection. The performance of detection filters obtained with 3rd-order (HOSD3) and 4th-order HOS (HOSD4) higher-order spectral decompositions, using the iterative method of Section 5.1 compared with two implementations of kurtosis-maximizing blind deconvolution, Deflation [13] and Med2d [43], with additive Gaussian noise, following the procedure in Fig. 4. **A**: Area under the curve at multiple in- and out-band SNRs for each method (*rows 1–4*) and the difference of HOSD3 and Deflation (*row 4*) at different orders in the deconvolving filter (*columns*). Both HOSD methods failed at the lowest order (16 taps); this we attribute to the use of tapered windows in spectral and filter estimation, which restricted the observable timescale beyond the limit imposed by filter order. Relative performance of both HOSD methods increased with increasing filter order, with pronounced superiority at low out-band SNRs ($< -15$ dB). The performance of Deflation and Med2d diminished at orders greater than 64 and at low SNRs, in part due to a tendency to converge on degenerate solutions, with the filter reproducing segments of the data. **B**: Average processing time required by each algorithm across filter orders. HOSD3 and Med2d performed comparably, with HOSD3 slightly faster at orders less than 128 and Med2d slightly faster at 128 and 256. Deflation lagged behind the first two, while HOSD4 was slowest, overall, reflecting a cubic increase in the number of values in the 4th-order HOS estimate with filter order.

*10.1. ECG recovery from Gaussian and non-Gaussian noise*

To show that these properties translate to settings more relevant to the real world, a similar analysis was conducted on two-minute segments of ECG recording obtained from the MIT-BIH database [27]. Results summarized in Fig. 3 demonstrate the ability of the algorithm to recover an averaged ECG waveform from extremely noisy data, according with what was observed for the artificial test case.

To demonstrate the application to signals corrupted by non-Gaussian noise, a similar test was repeated with the same ECG set using additive non-Gaussian noise. Noise in this case was generated by applying FIR filters with the same in- and out-band amplitude responses as in the Gaussian test to squared Gaussian white noise, whose distribution is $\chi^2(1)$. Because HOS preserve information about the phase response of the filter, it is also necessary to randomize the filter phase to avoid making the test overly specific to some particular (e.g. zero-phase) filter. The phase response was therefore randomized, but also smoothed to preserve the approximate time window of the zero-phase filter, so that deterministic filter HOS remained smooth at a similar scale as that of the signal.

Results for both Gaussian and non-Gaussian noise are shown in Fig. 4 and compared to the performance of a representative non-blind wavelet filter tailored to ECG detection. Comparison between signal recovery in the first recovered component vs the first four in Fig. 4 shows that the signal was often recovered among components 2–4 when the noise spectrum was dominant in the first component. As expected, the performance of the algorithm is degraded compared to the case of Gaussian noise, but it continues to fare well next to the wavelet filter, performing comparably at most signal-to-noise levels.

*10.2. Decomposition of multi-component signals*

The same approach applied in the previous section to signal recovery from non-Gaussian noise allows for the mutual separation of transient components within multi-component signals. Its application is demonstrated here through the blind classification of normal and abnormal heartbeats in a 10 minute sample of ECG from the MIT-BIH Arrhythmia Database [27] (Fig. 5). Reflecting the similarities of the respective waveforms, bispectral components do not map exactly onto heartbeat types. However, a projection into a low-dimensional feature space is obtained through loadings onto components at peak times. Waveforms may be blindly separated within this space using standard clustering techniques, such as k-means clustering. Using this approach, blind heartbeat detection and classification closely matched clinical annotation by a human observer (Fig. 5C), provided with the example data, set for the three most prevalent beat types.

*10.2.1. Comparison with MED filter estimation*

As reviewed in Section 7, the filter estimation step in (92) is qualitatively similar to moment-based techniques of blind deconvolution. We therefore compared two publicly available implementations of kurtosis maximizing blind deconvolution, Med2d [45] and Deflation [15] with HOS decomposition using the bispectrum (HOSD3) and trispectrum (HOSD4), in heartbeat detection. The results are summarized in Fig. 6. The four algorithms produced similar results at intermediate filter orders and noise levels, with some noteworthy differences elsewhere. First, both HOSD algorithms failed at the lowest filter order, 16 taps, which we attribute to the restriction of the time scale of the filter window, beyond that imposed by filter order alone, from the use of tapered windows in the averaged HOS and partial delay estimators. Second, HOSD outperformed the other methods in heartbeat detection with increasing filter order and decreasing out-band SNRs: while performance of MED estimation degrading rapidly between $-10\,\text{dB}$ and $-15\,\text{dB}$ out-band SNR, both HOSD algorithms performed above chance beyond SNRs of $-20\,\text{dB}$ at filter orders greater than 64. This effect is attributed in part to a tendency for the MED algorithms to converge on degenerate solutions at high filter orders and low SNRs when the filter order approached the square root of the total signal duration. Such solutions reproduce segments of the input data in the filter, which results in a single large impulse (and thus high kurtosis) [44]. Because the HOSD algorithm operates on averaged HOS estimators rather than the entire signal at once, it is less susceptible to this form of degeneracy, while robustness to out-band noise is aided by the noise optimization within HOS window design.

Finally, processing time and its dependence on filter order differed across the four algorithms (Fig. 6B): HOSD3 outperformed the other algorithms at filter orders less than 128, while Med2d performed with comparable speed at 128 and 256 taps. Deflation lagged behind the first two algorithms, while HOSD4 was slowest overall, with a more pronounced dependence on filter order. While computational complexity in both MED algorithms and HOSD3 depended quadratically on the filter order, the dependence is cubic for HOSD4, due to the addition of a frequency dimension in the trispectrum. This fact is reflected in the more rapid growth in processing time for HOSD4.

## 11. Discussion and conclusion

We have described a new method for obtaining noise-optimized filters from both deterministic and non-deterministic higher-orders spectra with no foreknowledge of underlying signal waveforms, assuming only their stable recurrence across records. In the non-deterministic case, the method yields a decomposition of HOS into approximately deterministic spectra. Algorithm 1 summarizes the prototypical form for the algorithm, as described in Section 8.1. We have reviewed the similarities between the HOSD algorithm and prior techniques of blind deconvolution based on moment maximization. The two procedures are related by way of Lemma 1, which shows that moment maximization produces essentially a

**Algorithm 1** HOS Decomposition.

1. Filter estimation
   (a) From a collection of $L$ records, $\{x_{0j}\}$, obtain an estimate of the $K$th-order HOS, $M_0^K$, and normalization, $D_0$, and compute the HOS weighting filter as $H_0 = M_0^{K*}/D_0$.
   (b) Obtain the average partial delay estimation filter, $G_0^{(0)} = \frac{1}{L}\sum_{j=1}^L G_{0j}$ where
   
   $G_{0j}(\omega_1) = \int X_{0j}(\omega_2)\ldots X_{0j}^*(\omega_K)H_0(\omega_1,\ldots,\omega_2)\delta(\omega_1 + \cdots + \omega_K)$
   
   $d\omega_2\ldots d\omega_K$
   
   (c) Apply the average delay filter to each record and take the maximum as the delay estimate:
   
   $\hat{\tau}_{0j}^{(0)} = \arg\max_t g_0^{(0)} * x_j(t)$
   
   (d) Reestimate the filter after delay compensation, and repeat through M steps until stopping criteria are met.
   
   $G_0^{(m+1)} = \frac{1}{L}\sum_{j=1}^L e^{i\omega\hat{\tau}_j^{(m)}} G_{0j}$

2. Feature estimation.
   (a) Following filter estimation, the feature waveform may be estimated as
   
   $\hat{F}_0(\omega) = \frac{1}{L}\sum_{j=1}^L \text{sgn}[g_0 * x_{0j}(\hat{\tau}_{0j})] e^{i\omega\hat{\tau}_{0j}^{(M)}} X_j(\omega)$

3. Component signal estimation
   (a) Threshold selection: choose a threshold, $\theta$, such that the $K$th-order zero-lag cumulant of the subthreshold filtered signal vanishes (or else attains some other statistically motivated value):
   
   $\theta : \frac{1}{L}\sum_j \text{cum}\{(g_0 * x_{0j}) - w_\theta[g_0 * x]\} = 0$
   
   where $w$ applies the thresholding, $w_\theta[z] = s(z - \theta)z$, for hard or soft threshold function, $s$.
   (b) Recover the component signal by convolving the waveform with the thresholded filter output,
   
   $\hat{y}_{0j} = af * w_\theta[g_0 * x]$
   
   where the scaling, $a$, is chosen to minimize mean squared error over records,
   
   $\hat{a} = \arg\min_a \sum \|x_{0j} - \hat{y}_{0j}\|^2$

4. Deflation
   (a) Subtract component estimates $\{x_{(p+1)j}\} = \{x_{pj} - \hat{y}_{pj}\}$
   (b) Repeat from the beginning on the residual: $\{x_{(p+1)j}\} \to \{x_{0j}\}$, go to step 1a to obtain $G_{(p+1)}$, $\hat{F}_{(p+1)}$, and $\{\hat{y}_{(p+1)j}\}$ matched filter, when applied to transient deterministic signals. Though we have motivated and developed the algorithm by way of the detection problem, Lemma 1 establishes its relevance to blind deconvolution more broadly.

In comparing it to MED, we identified some advantages of the HOSD algorithm: (1) the steps in the algorithm of Section 8.1 involve element-wise addition, multiplication and averaging, requiring no matrix inversions, which makes the algorithm numerically stable and easy to adapt to real-time settings. (2) The algorithm recovers optimal filters directly from HOS estimates rather than through a gradient ascent, making it less prone to converge on local optima and less sensitive to initialization. (3) Because the algorithm operates on smoothed HOS estimators, rather than the entire signal at once, it is less prone than MED to converge on filters which generate a single large impulse by reproducing a segment of the input. (4) By windowing in the HOS domain, the technique introduces additional tools to improve the robustness of the algorithm, as well as new ways to obtain cumulant-like measures from moment spectra. These qualities are reflected especially in its superior performance at low out-band SNR and high filter orders in the ECG test data (Fig. 6).

Conventional MED, by itself, also does not provide any basis for separating component signals. As described in Section 6, the application of a signal-dependent time window to the output is an important step in obtaining such a decomposition. To justify the use of a simple threshold window, reminiscent of windowing commonly found in wavelet filters [25], we considered how the thresholded output approximates the distribution of feature occurrence times (in Section 6.2), and how the residual roughly orthogonalizes HOS (in Section 6.3).

Some possible drawbacks of the present technique should be noted as well. First, the FIR filter estimation step underperformed standard MED at low filter orders; we attribute this effect to added restrictions of time scale imposed by tapered windows used in HOS estimation. Second, while computational demands of HOSD3 are comparable to MED, the HOSD3 algorithm implicitly maximizes skewness, thus it is not sensitive to features that are symmetrically distributed with respect to sign and is otherwise insensitive to narrowband signal features [20]. For these reasons, HOSD4 may be more suited to some contexts, but it is evident in Fig. 6B, that its cubic dependence on filter order puts HOSD4 at a computational disadvantage relative to conventional MED. Windowing in the HOS domain allows for some flexibility in the design of estimators, and it may be possible to address these shortcomings by restricting the underlying estimators to subdomains relevant for some particular signal of interest, thereby reducing computational overhead. Because one can relate such windowed estimators to semi-inner products in HOS space, as described in Section 2.2, matched filters are generally still contained in the space of filters which maximize the related zero-lag measure. Windowed estimators thus retain their essential relationship to the underlying optimization problem. A more detailed consideration of such extensions we leave for the future.

**Conflict of interest**

Technology related to this work is the subject of U.S. provisional patent application No. 62/695,586 and international patent application PCT/IB2019/055678.

**Acknowledgement**

Funded by National Institutes of Health 5R01DC004290-18. We wish to thank Alexander Billig, Phillip Gander, and Hiroyuki Oya.

**Appendix A. Development for multivariate and complex signals**

With the aid of the background given in Sections 2.1.5 and 2.1.6 and Corollary 1.5, development of the algorithm for the univariate real case may be directly generalized to both multivariate and complex signals, as explained next.

*A1. Multivariate signals*

For a multivariate signal, $x(t) \in \mathbb{R}^m$, our model supposes that distinct transient features appear within the component signals with fixed relative timing in a background of correlated Gaussian noise:

$$x(t) = [x_1(t), \ldots, x_m(t)] = f(t-\tau) + n(t-\tau)$$
$$= [f_1(t-\tau), \ldots, f_m(t-\tau)] + [n_1(t-\tau), \ldots, n_m(t)] \quad (A.1)$$

where $f(t) \in \mathbb{R}^m$ is the set of features appearing in the respective components and $n(t) \in \mathbb{R}^m$ is Gaussian noise with cross-spectral density $\mathrm{E}[N^*(\omega)N^T(\omega)] = \Sigma_n(\omega) = [\sigma_{ij}(\omega)]_{m \times m}$. We wish to find a set of filters $h = [h_1, \ldots, h_m]$, whose combined output,

$$r(t) = \sum_{j=1}^{m} h_j * x_j(t) \quad (A.2)$$

yields a minimally perturbed peak at $\tau$.

The argument is developed in the same way as in Section 3.1. In place of Eq. (31) we have

$$\mathrm{E}[r''(\tau)] = \int \omega^2 \sum_{i=1}^{m} F_i(\omega) H_i(\omega) \, d\omega = \int \omega^2 F^T H(\omega) \, d\omega \quad (A.3)$$

while the presence of correlated noise implies that (34) becomes

$$\mathrm{E}\left[(r'(\tau))^2\right] = \int \omega^2 \sum_{i=1, j=1}^{m} \sigma_{ij}(\omega) H_i(\omega) H_j^*(\omega) \, d\omega$$
$$= \int \omega^2 H^T \Sigma H^*(\omega) \, d\omega \quad (A.4)$$

The multivariate matched filter is found by completing the square within the Lagrangian function (35), as was done in (36):

$$H(\omega) = \Sigma^{-1} F^*(\omega) \quad (A.5)$$

for $\omega \neq 0$.

In applying the argument to the bispectral delay estimator, (51) generalizes as

$$r_{122}(t) = \sum_{\{j_1,j_2,j_3\}} \int X_{1j_1}(\omega_1) e^{i\omega_1 t} \int X_{2j_2}(\omega_2) X_{2j_3}^*(\omega_1+\omega_2)$$
$$\times H_{j_1 j_2 j_3}(\omega_1, \omega_2) \, d\omega_2 \, d\omega_1 \quad (A.6)$$

whose expected squared derivative, following (54), becomes

$$\mathrm{E}\left[(r'_{122}(\Delta\tau_{12}))^2\right] \leq A^2 \int \omega_1^2 \sum_{\{i_1,i_2,i_3\}, \{j_1,j_2,j_3\}}$$
$$\times \mathrm{E}[XX^*_{1i_1 j_1}(\omega_1) XX^*_{2i_2 j_2}(\omega_2) XX^*_{2i_3 j_3}(\omega_1+\omega_2)]$$
$$\times H^*_{i_1 i_2 i_3} H_{j_1 j_2 j_3}(\omega_1, \omega_2) \, d\omega_2 \, d\omega_1 \quad (A.7)$$

where $XX^*_{abc}(\omega)$ is shorthand for $X_{ab}(\omega) X^*_{ac}(\omega)$.

Eq. (52) generalizes as

$$\mathrm{E}[r''_{122}(\Delta\tau_{12})] = -\int \omega_1^2 \sum_{\{j_1,j_2,j_3\}} F_{j_1}(\omega_1) \int F_{j_2}(\omega_2) F^*_{j_3}(\omega_1+\omega_2)$$
$$\times H_{j_1 j_2 j_3}(\omega_1, \omega_2) \, d\omega_2 \, d\omega_1 \quad (A.8)$$

The Lagrangian (35) reduces, as before, to an easily solved quadratic form, which is minimized at

$$\mathbf{H}(\omega_1, \omega_2) = \mathbf{D}^{-1}(\omega_1, \omega_2)[F(\omega_1) \otimes F(\omega_2) \otimes F^*(\omega_1+\omega_2)] \quad (A.9)$$

where $\otimes$ is the Kronecker tensor product, $\mathbf{H} = [H_{\{j_1,j_2,j_3\}}]_{m^3 \times 1}$, and

$$\mathbf{D} = \mathrm{E}\left[[XX^*_{1i_1 j_1}(\omega_1) XX^*_{2i_2 j_2}(\omega_2) XX^*_{2i_3 j_3}(\omega_1+\omega_2)]_{m^3 \times m^3}\right] \quad (A.10)$$

or in the case when $n$ is strictly Gaussian noise, with cross-spectral density matrix $\Sigma(\omega)$:

$$\mathbf{D}(\omega_1, \omega_2) = \Sigma(\omega_1) \otimes \Sigma(\omega_2) \otimes \Sigma(\omega_1+\omega_2) \quad (A.11)$$

where $\Sigma(\omega) = \Sigma_n(\omega) + F^*F^T(\omega)$. These results generalize directly to HOS of arbitrary order, $K$, with

$$\mathbf{H}_{m^K \times 1}(\omega_1, \ldots, \omega_{K-1}) = \mathbf{D}_{m^K \times m^K}^{-1}(\omega_1, \ldots, \omega_{K-1})$$
$$\times \left[ F(\omega_1) \otimes \cdots \otimes F(\omega_{K-1}) \otimes F^*\left(\sum_{k=1}^{K-1} \omega_k\right) \right] \quad (A.12)$$

and $\mathbf{D}(\omega_1, \ldots, \omega_{K-1}) = \Sigma(\omega_1) \otimes \cdots \otimes \Sigma(\omega_{K-1}) \otimes \Sigma(\sum_{k=1}^{K-1} \omega_k)$ for Gaussian noise.

### A2. Multivariate filter estimation

Filter estimation follows from a similar generalization of the univariate algorithm. The complete set of $\binom{K+m-1}{m-1}$ higher order auto- and cross-spectral estimators, $\hat{\mathbf{M}}$, provide an estimate of $[F(\omega_1) \otimes F(\omega_2) \ldots \otimes F^*(\omega_1 + \cdots + \omega_{K-1})]$, and $\mathbf{D}$ may be estimated from multivariate generalizations of polycoherence:

$$\hat{D}^{\{i_1,\ldots,i_K\}}_{\{j_1,\ldots,j_K\}} = \sigma_{i_1 j_1}(\omega_1) \langle XX^*_{i_2 j_2}(\omega_2) \ldots XX^*_{i_K j_K}(\omega_1 + \cdots + \omega_{K-1}) \rangle \quad (A.13)$$

or under the assumption of Gaussian noise as

$$\hat{D}^{\{i_1,\ldots,i_K\}}_{\{j_1,\ldots,j_K\}} = \sigma_{i_1 j_1}(\omega_1) \sigma_{i_2 j_2}(\omega_2) \ldots \sigma_{i_K j_K}(\omega_1 + \cdots + \omega_{K-1}) \quad (A.14)$$

In the case of the bispectrum ($K=3$), the following $\frac{1}{6}(m+2)(m+1)m$ spectra describe the complete set of $3^{\text{rd}}$-order interactions:

$$\hat{M}_{\{j_1, j_2, j_3\}}(\omega_1, \omega_2) \to F_{j_1}(\omega_1) F_{j_2}(\omega_2) F^*_{j_3}(\omega_1 + \omega_2) \quad (A.15)$$

and with multivariate bicoherence normalization

$$\hat{D}^{\{i_1, i_2, i_3\}}_{\{j_1, j_2, j_3\}} = \sigma_{i_1 j_1}(\omega_1) \langle XX^*_{i_2 j_2}(\omega_2) XX^*_{i_3 j_3}(\omega_1 + \omega_2) \rangle \quad (A.16)$$

$\hat{\mathbf{H}}$ is obtained by substituting $\hat{\mathbf{D}}$ and $\hat{\mathbf{M}}$ into Eq. (A.12).

Next, a detection filter is obtained for the $i$th component through a summation of Eq. (92) over all spectra in which $i$ is the first component:

$$G_i^{(\nu+1)} = \frac{1}{L} \sum_{l=1}^{L} e^{i\omega_1 \hat{\tau}_l^{(\nu)}} \sum_{j=1, k=1}^{m} \sum_{\omega_2=-W}^{W} X_{lj}[\omega_2] X^*_{lk}[\omega_1 + \omega_2]$$
$$\times \hat{H}_{ijk}[\omega_1, \omega_2] \quad (A.17)$$

Finally, the delay is estimated for the $l$th record by summing the respective filter outputs over all components, and the algorithm proceeds to the next iteration:

$$\hat{\tau}_l^{(\nu+1)}(t) = \hat{\tau}_l^{(\nu)} + \arg\max_t \left[ \sum_{j=1}^{m} g_j^{(\nu)} * x_j(t + \hat{\tau}_l^{(\nu)}) \right] \quad (A.18)$$

Note that the delay, $\tau_l$, is necessarily assumed to be the same across all component signals in the $l$th record, but any fixed relative offset of delays between components, which does not change across records, will be reflected in the linear phase of the respective filters, thus posing no limitation.

The computational complexity of estimating all non-redundant spectra grows as $m^3$ with the number of variates ($m^K$ generally). This cubic growth of complexity may be addressed by including only a subset of spectra. For example, if noise is uncorrelated across components, then $\mathbf{D}^{-1}$ is asymptotically diagonal with decreasing signal to noise ratio (exactly so for the noise-only form of $\mathbf{D}$), thus it is appropriate to include only the $m$ auto-spectra in filter estimation. Otherwise one might restrict estimation to the $m^2$ auto- and pairwise cross-spectral interactions, which is adequate to account for the correlation structure of Gaussian noise.

### A3. Complex signals

Using the isomorphism described in Section 2.1.6 between the HOS of $M^{\{i_1, i_2, i_3\}}[y]$, for $y(t) \in \mathbb{C}^1$, and the bivariate spectrum of $M_{\{j_1, j_2, j_3\}}[x_+, x_-]$, for $[x_+, x_-] \in \mathbb{R}^2$, the extension to a complex univariate signal, $y$, may be developed from the bivariate case following from A.1, where the respective component signals are obtained from the real and imaginary parts of $y$ given in Eq. (13)

$$x_+(t) = \frac{1}{2}(\text{Re}\{y\} - \mathcal{H}\{\text{Im}\{y\}\})$$
$$x_-(t) = \frac{1}{2}(\text{Re}\{y\} + \mathcal{H}\{\text{Im}\{y\}\}) \quad (A.19)$$

By way of Eq. (15), non-redundant symmetry regions may be distinguished by the signs of the three frequency arguments, ordered by frequency magnitude [17]. The resultant mapping is illustrated for the $M^{\{1,1,-1\}}$ spectrum in Fig. A.1.

| Region | $\|\omega_a\| \leq$ $j_a$ | $\|\omega_b\| \leq$ $j_b$ | $\|\omega_c\|$ $j_c$ | Count |
|--------|---|---|---|---|
| $A^+$ | + | + | + | 2 |
| $A^-$ | − | − | − | 2 |
| $C_1^+$ | − | + | + | 2 |
| $C_1^-$ | + | − | − | 2 |
| $C_2^+$ | + | − | + | 2 |
| $C_2^-$ | − | + | − | 2 |
| $C_3^+$ | + | + | − | 0 |
| $C_3^-$ | − | − | + | 0 |

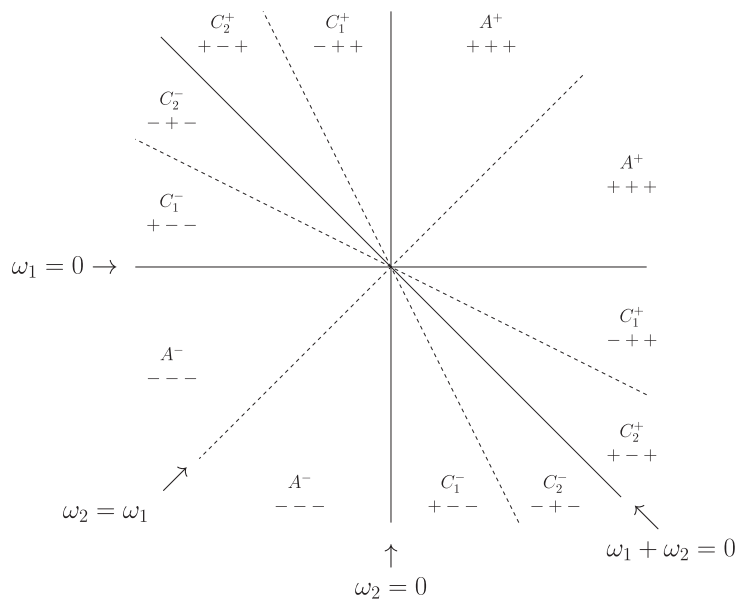

**Fig. A.1.** Symmetry regions within $M^{\{1,1,-1\}}[y]$ and their mapping to $M_{\{j_1, j_2, j_3\}}[x_+, x_-]$. Lines indicate planes of reflection about $w_a = 0$ (*solid lines*) and $w_a = w_b$ (*dashed lines*). The $\{1, 1, -1\}$ spectrum is symmetric under permutation of the first and second terms, hence about the line $\omega_2 = \omega_1$. Symmetry regions are distinguished by the signs of the frequency arguments ordered by frequency magnitude.

In practice, it may be advantageous to compute the spectrum directly from $y$, rather than through an explicit transformation to the bivariate form. Spectra for which $i_1 + i_2 + i_3 = \pm 1$ include 6 of the 8 non-redundant symmetry regions, and are sufficient to uniquely recover a deterministic transient signal (see Corollary 1.5). However, differences in how the symmetry regions are sampled, in practice, imply that they do not give precisely equivalent information on the signal. while the two neglected regions, $C_3^+$ and $C_3^-$, in Fig. A.1, are contained only in the $\{1, 1, 1\}$ and $\{-1, -1, -1\}$ spectra.

# References

[1] S. Alshebeili, A. Cetin, A phase reconstruction algorithm from bispectrum (seismic reflection data), IEEE Trans. Geosci. Remote Sens. 28 (2) (1990) 166–170.
[2] P.O. Amblard, M. Gaeta, J.L. Lacoume, Statistics for complex variables and signals–part II: signals, Signal Process. 53 (1) (1996) 15–25.
[3] H. Bartelt, A.W. Lohmann, B. Wirnitzer, Phase and amplitude recovery from bispectra, Appl. Opt. 23 (18) (1984) 3121–3129.
[4] M. Bartlett, The spectral analysis of point processes, J. R. Stat. Soc. 25 (2) (1963) 264–296.
[5] A.J. Bell, T.J. Sejnowski, An information-maximization approach to blind separation and blind deconvolution, Neural Comput. 7 (6) (1995) 1129–1159.
[6] T. Bendory, N. Boumal, C. Ma, Z. Zhao, A. Singer, Bispectrum inversion with application to multireference alignment, IEEE Trans. Signal Process. 66 (4) (2018) 1037–1050.
[7] M. Boussé, O. Debals, L. De Lathauwer, Tensor-based large-scale blind system identification using segmentation, IEEE Trans. Signal Process. 65 (21) (2017) 5770–5784.
[8] D.R. Brillinger, An introduction to polyspectra, Ann. Math. Stat. 36 (5) (1965) 1351–1374.
[9] D.R. Brillinger, M. Rosenblatt, Asymptotic theory of estimates of kth-order spectra, Proc. Natl. Acad. Sci. USA 57 (2) (1967) 206.
[10] M.K. Broadhead, L.A. Pflug, R.L. Field, Use of higher order statistics in source signature estimation, J. Acoust. Soc. Am. 107 (5) (2000) 2576–2585.
[11] J.A. Cadzow, Blind deconvolution via cumulant extrema, IEEE Signal Process. Mag. 13 (3) (1996) 24–42.
[12] G.C. Carter, Coherence and time delay estimation, Proc. IEEE 75 (2) (1987) 236–255.
[13] M. Castella, Matlab toolbox for separation of convolutive mixtures, Telecom SudParis, 2011. http://www-public.imtbs-tsp.eu/~castella/toolbox/.
[14] M. Castella, A. Chevreuil, J.C. Pesquet, Convolutive mixtures, in: Handbook of Blind Source Separation, Elsevier, 2010, pp. 281–324.
[15] M. Castella, E. Moreau, New kurtosis optimization schemes for miso equalization, IEEE Trans. Signal Process. 60 (3) (2012) 1319–1330.
[16] B. Castro, D. Kogan, A. Geva, ECG feature extraction using optimal mother wavelet, in: Electrical and Electronic Engineers in Israel, 2000. The 21st IEEE Convention of the IEEE, 2000, pp. 346–350.
[17] V. Chandran, S. Elgar, A general procedure for the derivation of principal domains of higher-order spectra, IEEE Trans. Signal Process. 42 (1) (1994) 229–233.
[18] B. Chen, A.P. Petropulu, Frequency domain blind mimo system identification based on second-and higher order statistics, IEEE Trans. Signal Process. 49 (8) (2001) 1677–1688.
[19] C.Y. Chi, C.H. Chen, Cumulant-based inverse filter criteria for mimo blind deconvolution: properties, algorithms, and application to DS/CDMA systems in multipath, IEEE Trans. Signal Process. 49 (7) (2001) 1282–1299.
[20] W. Collis, P. White, J. Hammond, Higher-order spectra: the bispectrum and trispectrum, Mech. Syst. Signal Process. 12 (3) (1998) 375–394.
[21] J.W. Dalle Molle, Higher-Order Spectral Analysis and the Trispectrum, University of Texas at Austin, 1992 Ph. d. Thesis.
[22] N. Delfosse, P. Loubaton, Adaptive blind separation of independent sources: a deflation approach, Signal Process. 45 (1) (1995) 59–83.
[23] I. Domanov, L. De Lathauwer, Blind channel identification of miso systems based on the CP decomposition of cumulant tensors, in: Signal Processing Conference, 2011 19th European, IEEE, 2011, pp. 2215–2218.
[24] D. Donoho, On minimum entropy deconvolution, in: Applied Time Series Analysis II, Elsevier, 1981, pp. 565–608.
[25] D.L. Donoho, I.M. Johnstone, Ideal denoising in an orthonormal basis chosen from a library of bases, Compt. Rendus Acad. Sci. 319 (12) (1994) 1317–1322.
[26] J. Fonollosa, C. Nikias, Wigner higher order moment spectra: definition, properties, computation and application to transient signal analysis, IEEE Trans. Signal Process. 41 (1) (1993) 245.
[27] A.L. Goldberger, L.A.N. Amaral, L. Glass, J.M. Hausdorff, P.C. Ivanov, R.G. Mark, J.E. Mietus, G.B. Moody, C.K. Peng, H.E. Stanley, Physiobank, physiotoolkit, and physionet, Circulation 101 (23) (2000) e215–e220.
[28] S. Hagihira, M. Takashina, T. Mori, T. Mashimo, I. Yoshiya, Practical issues in bispectral analysis of electroencephalographic signals, Anesth. Analg. 93 (4) (2001) 966–970.
[29] D. Hatzinakos, C.L. Nikias, Blind equalization using a tricepstrum-based algorithm, IEEE Trans. Commun. 39 (5) (1991) 669–682.
[30] M.J. Hinich, G.R. Wilson, Time delay estimation using the cross bispectrum, IEEE Trans. Signal Process. 40 (1) (1992) 106–113.
[31] A. Hyvarinen, Fast ICA for noisy data using gaussian moments, in: Circuits and Systems, 1999. ISCAS '99. Proceedings of the 1999 IEEE International Symposium on, vol. 5, 1999, pp. 57–61.
[32] M.G. Kang, K.T. Lay, A.K. Katsaggelos, Phase estimation using the bispectrum and its application to image restoration, Opt. Eng. 30 (7) (1991) 976–986.
[33] Y.C. Kim, E.J. Powers, Digital bispectral analysis and its applications to nonlinear wave interactions, IEEE Trans. Plasma Sci. 7 (2) (1979) 120–131.
[34] C.K. Kovach, A biased look at phase locking: brief critical review and proposed remedy, IEEE Trans. Signal Process. 65 (17) (2017) 4468–4480.
[35] C.K. Kovach, Implementation of Higher Order Spectral Decomposition, 2019, https://github.com/ckovach/HOSD.git.
[36] C.K. Kovach, P.E. Gander, The demodulated band transform, J. Neurosci. Methods 261 (2016) 135–154.
[37] C.K. Kovach, H. Oya, H. Kawasaki, The bispectrum and its relationship to phase-amplitude coupling, NeuroImage 173 (2018) 518–539.
[38] M. Kreutz, B. Völpel, H. Janßen, Scale-invariant image recognition based on higher-order autocorrelation features, Pattern Recognit. 29 (1) (1996) 19–26.
[39] X. Li, N.M. Bilgutay, Wiener filter realization for target detection using group delay statistics, IEEE Trans. Signal Process. 41 (6) (1993) 2067–2074.
[40] A.W. Lohmann, G. Weigelt, B. Wirnitzer, Speckle masking in astronomy: triple correlation theory and applications, Appl. Opt. 22 (24) (1983) 4028–4037.
[41] R. Marabini, J. Carazo, Practical issues on invariant image averaging using the bispectrum, Signal Process. 40 (2) (1994) 119–128.
[42] T. Matsuoka, T.J. Ulrych, Phase estimation using the bispectrum, Proc. IEEE 72 (10) (1984) 1403–1411.
[43] G.L. Mc Donald, Minimum entropy deconvolution multipack, Mathworks File Exchange (2015). https://www.mathworks.com/matlabcentral/fileexchange/53484-minimum-entropy-deconvolution-multipack-med-meda-omeda-momeda-mckd.
[44] G.L. Mc Donald, Q. Zhao, Multipoint optimal minimum entropy deconvolution and convolution fix: application to vibration fault detection, Mech. Syst. Signal Process. 82 (2017) 461–477.
[45] G.L. McDonald, Q. Zhao, M.J. Zuo, Maximum correlated kurtosis deconvolution and application on gear tooth chip fault detection, Mech. Syst. Signal Process. 33 (2012) 237–255.
[46] J.A. McLaughlin, J. Raviv, Nth-order autocorrelations in pattern recognition, Inf. Control 12 (2) (1968) 121–142.
[47] J.M. Mendel, Tutorial on higher-order statistics (spectra) in signal processing and system theory: theoretical results and some applications, Proc. IEEE 79 (3) (1991) 278–305.
[48] M. Nakamura, Waveform estimation from noisy signals with variable signal delay using bispectrum averaging, IEEE Trans. Biomed. Eng. 40 (2) (1993) 118–127.
[49] C.L. Nikias, J.M. Mendel, Signal processing with higher-order spectra, IEEE Signal Process. Mag. 10 (3) (1993) 10–37.
[50] C.L. Nikias, R. Pan, Time delay estimation in unknown gaussian spatially correlated noise, IEEE Trans. Acoust. Speech Signal Process. 36 (11) (1988) 1706–1714.
[51] A.K. Ovacıklı, P. Pääjärvi, J.P. LeBlanc, J.E. Carlson, Recovering periodic impulsive signals through skewness maximization, IEEE Trans. Signal Process. 64 (6) (2016) 1586–1596.
[52] P. Pääjärvi, J.P. LeBlanc, Online adaptive blind deconvolution based on third-order moments, IEEE Signal Process. Lett. 12 (12) (2005) 863–866.
[53] R. Pan, C. Nikias, Phase reconstruction in the trispectrum domain, IEEE Trans. Acoust. Speech Signal Process. 35 (6) (1987) 895–897.
[54] R. Pan, C.L. Nikias, The complex cepstrum of higher order cumulants and non-minimum phase system identification, IEEE Trans. Acoust. Speech Signal Process. 36 (2) (1988) 186–205.
[55] A.P. Petropulu, C.L. Nikias, Signal reconstruction from the phase of the bispectrum, IEEE Trans. Signal Process. 40 (3) (1992) 601–610.
[56] L.A. Pflug, G.E. Ioup, J.W. Ioup, K.H. Barnes, R.L. Field, G.H. Rayborn, Detection of oscillatory and impulsive transients using higher-order correlations and spectra, J. Acoust. Soc. Am. 91 (5) (1992) 2763–2776.
[57] L.A. Pflug, G.E. Ioup, J.W. Ioup, R.L. Field, Prediction of signal-to-noise ratio gain for passive higher-order correlation detection of energy transients, J. Acoust. Soc. Am. 98 (1) (1995) 248–260.
[58] L.A. Pflug, G.E. Ioup, J.W. Ioup, R.L. Field, J.H. Leclere, Time-delay estimation for deterministic transients using second- and higher-order correlations, J. Acoust. Soc. Am. 94 (3) (1993) 1385–1399.
[59] V. Popovici, J. Thiran, Higher order autocorrelations for pattern classification, in: Proceedings 2001 International Conference on Image Processing (Cat. No.01CH37205), vol. 3, 2001, pp. 724–727.
[60] B.M. Sadler, G.B. Giannakis, Shift-and rotation-invariant object reconstruction using the bispectrum, JOSA A 9 (1) (1992) 57–69.
[61] K. Sasaki, T. Sato, Y. Yamashita, Minimum bias windows for bispectral estimation, J. Sound Vib. 40 (1) (1975) 139–148.
[62] N.D. Sidiropoulos, L. De Lathauwer, X. Fu, K. Huang, E.E. Papalexakis, C. Faloutsos, Tensor decomposition for signal processing and machine learning, IEEE Trans. Signal Process. 65 (13) (2017) 3551–3582.
[63] C. Simon, P. Loubaton, C. Vignat, C. Jutten, G. d'Urso, Separation of a class of convolutive mixtures: a contrast function approach, in: Acoustics, Speech, and

Signal Processing, 1999. Proceedings., 1999 IEEE International Conference on, vol. 3, IEEE, 1999, pp. 1429–1432.

[64] A.V. Totsky, A.A. Zelensky, V.F. Kravchenko, Bispectral Methods of Signal Processing: Applications in Radar, Telecommunications and Digital Image Restoration, Walter de Gruyter GmbH & Co KG, 2015.

[65] J.K. Tugnait, Identification and deconvolution of multichannel linear non-Gaussian processes using higher order statistics and inverse filter criteria, IEEE Trans. Signal Process. 45 (3) (1997) 658–672.

[66] G. Turin, An introduction to matched filters, IRE Trans. Inf. Theory 6 (3) (1960) 311–329.

[67] R.A. Wiggins, Minimum entropy deconvolution, Geoexploration 16 (1) (1978) 21–35.

[68] C.D. Woody, Characterization of an adaptive filter for the analysis of variable latency neuroelectric signals, Med. Biol. Eng. 5 (6) (1967) 539–554.